\documentclass[twocolumn]{aastex631}

\usepackage{amsmath}
\usepackage[encapsulated]{CJK}
\usepackage{ucs}
\usepackage[utf8x]{inputenc}
\newcommand{\cntext}[1]{\begin{CJK}{UTF8}{gbsn}#1\ignorespacesafterend\end{CJK}}

\usepackage{bm}

\newcommand{\vel}{\mathrm{v}}

\def\bfnabla{{\mbox{\boldmath $\nabla$}}}

\newcommand\bvp{{\mbox{\boldmath $v$}}}

\newcommand\bP{{\mbox{\boldmath $P$}}}
\newcommand\bn{{\mbox{\boldmath $n$}}}

\newcommand\bF{{\mbox{\boldmath $F$}}}

\newcommand\Crat{{\mathbb{C}}}
\newcommand\Prat{{\mathbb{P}}}
\newcommand\Rrat{{\mathbb{R}}}
\def\<{\,\langle\langle}
\def\>{\,\rangle\rangle}

\begin{document}

\title{Thermal Structure Determines Kinematics: \\Vertical Shear Instability in Stellar Irradiated Protoplanetary Disks}

\correspondingauthor{Shangjia Zhang}
\email{zhangs17@unlv.nevada.edu}

\author[0000-0002-8537-9114]{Shangjia Zhang (\cntext{张尚嘉})}
\affiliation{Department of Physics and Astronomy, University of Nevada, Las Vegas, 4505 S. Maryland Pkwy, Las Vegas, NV, 89154, USA}
\affiliation{Nevada Center for Astrophysics, University of Nevada, Las Vegas, Las Vegas, NV 89154, USA}

\author[0000-0003-3616-6822]{Zhaohuan Zhu (\cntext{朱照寰})}
\affiliation{Department of Physics and Astronomy, University of Nevada, Las Vegas, 4505 S. Maryland Pkwy, Las Vegas, NV, 89154, USA}
\affiliation{Nevada Center for Astrophysics, University of Nevada, Las Vegas, Las Vegas, NV 89154, USA}

\author[0000-0002-2624-3399]{Yan-Fei Jiang (\cntext{姜燕飞})}
\affiliation{Center for Computational Astrophysics, Flatiron Institute, New York, NY 10010, USA}

\begin{abstract}

Turbulence is crucial for protoplanetary disk dynamics, and Vertical Shear Instability (VSI) is a promising mechanism in outer disk regions to generate turbulence. We use Athena++ radiation module to study VSI in full and transition disks, accounting for radiation transport and stellar irradiation. We find that the thermal structure and cooling timescale significantly influence VSI behavior. The inner rim location and radial optical depth affect disk kinematics. Compared with previous vertically-isothermal simulations, our full disk and transition disks with small cavities have a superheated atmosphere and cool midplane with long cooling timescales, which suppresses the corrugation mode and the associated meridional circulation. This temperature structure also produces a strong vertical shear at $\mathrm{\tau_*}$ = 1, producing an outgoing flow layer at $\tau_* < 1$ on top of an ingoing flow layer at $\tau_* \sim 1$. The midplane becomes less turbulent, while the surface becomes more turbulent with effective $\alpha$ reaching $\sim10^{-2}$ at $\tau_* \lesssim$1. This large surface stress drives significant surface accretion, producing substructures. Using temperature and cooling time measured/estimated from radiation-hydro simulations, we demonstrate that less computationally-intensive simulations incorporating simple orbital cooling can almost reproduce radiation-hydro results.
By generating synthetic images, we find that substructures are more pronounced in disks with larger cavities. The higher velocity dispersion at the gap edge could also slow particle settling. Both properties are consistent with recent Near-IR and ALMA observations. Our simulations predict that regions with significant temperature changes are accompanied by significant velocity changes, which can be tested by ALMA kinematics/chemistry observations.

\end{abstract}


\section{Introduction}

Turbulence in protoplanetary disks plays significant roles in planet formation, such as determining mass accretion, angular momentum transport, and dust dynamics. In the disk midplane beyond 0.1 au, where the magnetorotational instability (MRI) is suppressed due to low ionization rates, turbulence can be generated by hydrodynamic instabilities \citep{turner14, armitage20, lesur22}. One promising candidate among these instabilities is the vertical shear instability (VSI) \citep{nelson13, stoll14, barker15, umurhan16a, lesur22}. The VSI is driven by the vertical differential rotation of the disk (dv$_{\phi}$/dZ $\neq$0). It occurs in baroclinic disks, characterized by non-parallel density and pressure gradients \citep{lesur22, klahr23}. Such a configuration is often satisfied in stellar irradiated protoplanetary disks, where the temperature decreases away from the central star in the radial direction. Most studies on the VSI assume a vertically constant temperature, which is thought to be a valid assumption near the disk midplane (e.g., \citealt{calvet91, chiang97}). 

A distinctive feature of the VSI is the presence of the corrugation mode in the meridional plane, which involves large-scale gas circulations in the vertical direction, while remaining confined in the radial direction \citep{nelson13, lyra19}. As a result, anisotropic turbulence arises in the Z-$\phi$ and R-$\phi$ stresses, with the former typically reaching magnitudes of $10^{-2}$ and the latter ranging from $10^{-4}$ to $10^{-3}$ \citep{stoll14, stoll16, stoll17a, flock17, manger18, manger20, manger21, pfeil21}. The amplitude of turbulence increases with the aspect ratio (i.e., temperature) of the disk \citep{manger20, manger21}. Simulations with mm-sized dust particles show that they can be stirred up very high above the midplane due to the strong turbulence in the vertical direction \citep{stoll17a, flock17, flock20, blanco21, dullemond22}. High resolution simulations show that strong shears between neighbouring bands of these meridional circulations can generate vortices in the R-Z plane \citep{flores-rivera20, klahr23, melonfuksman23b}. Three-dimensional simulations also demonstrate they can generate vortices and zonal flows in the R-$\phi$ plane due to the Kelvin-Helmholtz instability \citep{flock17, manger18, flock20, blanco21, pfeil21}. These zonal flows can possibly explain some of the ubiquitous substructures observed by ALMA dust continuum observations \citep[e.g., ][]{andrews18, long18, vandermarel19, cieza21, blanco21, andrews20, bae22}. 

The role of thermodynamics is crucial in determining whether the vertical shear instability (VSI) can occur \citep{lin15, lyra19, lesur22}. Specifically, the global mode of VSI requires a fast cooling timescale, where the cooling time normalized to the orbital time should be less than a threshold \citep{lin15}, 
\begin{equation}
\beta < \beta_c \equiv \Omega_K^{-1}(h/r)|q|/(\gamma_g - 1),
\label{eq:beta_c}
\end{equation}
where $\Omega_K$ is the Keplerian frequency, $h/r$ is the disk aspect ratio, $q$ is the radial temperature power-law index, and $\gamma_g$ is the adiabatic index. Analytical studies considering dust-gas coupling and dust evolution have identified regions where the VSI can operate \citep{malygin17, pfeil19, fukuhara21}. Disks with globally uniform cooling times larger than this critical cooling time do not develop VSI \citep{manger21}. However, short length-scale perturbations may still grow after evolving for a very long time \citep{klahr23, pfeil23}. In cases where the midplane has a long cooling timescale, yet the atmosphere has a short cooling time, studies using vertically isothermal simulations demonstrate the persistence of the VSI in the disk atmosphere \citep{pfeil21, fukuhara23}. The VSI in the unstable disk atmosphere can even penetrate the stable midplane, as long as the VSI-stable layer is less than two gas
scale heights and the VSI-unstable layer is thicker than two gas scale heights \citep{fukuhara23}. Additionally, with equivalent short cooling times, radiation hydrodynamic simulations produce similar results to vertically isothermal simulations \citep{stoll14, flock17}.

While VSI aligns with certain observational facets, such as a low level of R-$\phi$ turbulence, it also has tensions with some other facets.

\begin{itemize}
  \item \textit{A possible mechanism for substructures.} The (sub-)mm dust continuum emission traces $\sim$0.1-10 mm dust particles residing in the disk midplane, where substructures have been detected in a majority of observed disks when sufficient resolution is achieved \citep{bae22}. These substructures predominantly manifest as gaps and rings, although occasional arcs and spirals have been observed as well. VSI can generate density perturbations and vortices owing to the zonal flow led by the corrugation mode. However, these perturbation are typically small and requires very high resolution and sensitivity to detect them \citep{blanco21}.
  
  \item \textit{Consistent weak turbulence $\alpha_\mathrm{R}$.} The VSI generates a low level of turbulence in the R-$\phi$ plane \citep[$\alpha_\mathrm{R}$ $\sim$ 10$^{-5}$ - 10$^{-3}$,][]{nelson13, stoll14, flock17, lesur22}. This is consistent with planet-disk interactions models with ad hoc turbulent viscosity to explain the observed multiple gaps and rings \citep{dong17, bae17, zhang18, paardekooper22}. A low level of viscosity (10$^{-4}$ - 10$^{-3}$) is also needed to match the disk dispersal timescale ($\sim$ Myrs) in disk evolution models \citep{mulders17, lodato17, tabone22, manara22}.
  
  \item \textit{Overpredicted turbulence $\alpha_\mathrm{Z}$.} In the disk's vertical direction, edge-on and inclined disks exhibit remarkably thin dust emission layers, which can be translated to a low value of $\alpha_Z$/St \citep[$<$ 10$^{-2}$, ][]{pinte16, doi21, villenave20, villenave22, sierra20, ueda20, ueda21}, where $\alpha_Z$ is the turbulence in the Z-$\mathrm{\phi}$ plane, and St is the Stokes number that characterizes the gas-dust coupling. An exception is that the inner ring of HD 163296, which has $\alpha_Z$/St $>$ 1 \citep[]{doi21}. Adopting a typical Stokes number, the $\alpha_Z$ is estimated to be $\lesssim$ 10$^{-4}$ in most cases. However, VSI generates very strong turbulence in the vertical direction \citep[$\sim$10$^{-2}$, e.g., ][]{stoll14}, with the correlation at neighbouring radii, which conflicts with many observations \citep{dullemond22}.

  \item \textit{Yet-to-be-detected VSI corrugation mode.} Gas line emission observations demand higher sensitivities, yet data have been accumulating from recent and ongoing ALMA large programs such as, MAPS \citep{oberg21b}. With the aid of channel maps derived from these observations, we can measure the 3D velocity and temperature structure of the disk for a large sample of disks \citep[e.g., ][]{miotello22, pinte22}. \citet{barraza-alfaro21} predicts that the alternating blueshifted-redshifted corrugation mode can be observed in the CO channel maps given very high spectral resolution (50 m s$^{-1}$ at ALMA band 6). However, there is no firm detection of this pattern so far.
  
\end{itemize}

While disk thermodynamics plays a key role in VSI, most previous studies focused on simple thermodynamics such as locally isothermal or orbital cooling treatments. The vertical thermal structure is also underexplored. With a more self-consistent treatment, we can provide a more robust model and improve connections between observations and theory. 

To that end, we employ a self-consistent radiation-hydrodynamics approach with temperature-dependent DSHARP opacity \citep{birnstiel18}. Unlike previous simulations that utilized locally isothermal equations of state or flux-limited diffusion approximation with constant opacity, we utilize the Athena++ \citep{stone20} implicit radiation module \citep{jiang14, jiang21}. This module incorporates angle-dependent radiative transfer equations with implicit solvers to accurately model the disk radiation transport. The module can capture both optically thin and thick regimes and shadowing and beam crossing accurately. Additionally, we incorporate stellar irradiation using long-characteristic ray tracing as a heating source.

Recently, \citet{melonfuksman23a, melonfuksman23b} independently study VSI in irradiated protoplanetary disks using M1 method \citep{melonfuksman21}. While we focus on the outer disk beyond 20 au and they focus on 4-7 au in the inner disk, the results for our fiducial model are consistent with their dust depleted disk models, which show quiescent midplane and turbulent atmosphere.

The paper layout is the following. In Section~\ref{sec:methods}, we discuss the numerical setup of the simulations. In Section~\ref{sec:results}, we present results of radiation hydrodynamic simulations, accompanied with a series of pure hydro simulations. In Section~\ref{sec:discussion}, we discuss the formation of substructures, and observational/modeling prospect. Finally, we conclude the paper in Section~\ref{sec:conclusions}.

\section{Methods}
\label{sec:methods}

\subsection{Disk Model Setup}
\label{sec:disksetup}

In our study, we explored both full disks and transition disks with varying inner disk truncation radii or cavity sizes, denoted as $\rm r_{cav}$. 
Our simulations were performed in spherical polar coordinates $\rm {r, \theta, \phi}$, while most of the disk structure was set up in cylindrical coordinates $\rm {R, Z, \phi}$.

\begin{table*}
	\centering
	\caption{Simulation setups and measured values at later times. $h/r$ and $q$ are measured from midplane temperature in Figure \ref{fig:1DradialT}.}
    \begin{tabular}{ccccccccc}
        \hline
        model name & radiation & r$_{\mathrm{cav}}$  & (h/r)$_0$ & (h/r)$_0$ measured & $p$ & $p$ measured & $q$ & $q$ measured \\
        \hline
        \texttt{3r$_\odot$-rad} & on & 3r$_\odot$ & 0.07 & 0.049 & -2.25 & -2.5 & -0.5 & -0.2\\
        \texttt{18au-rad} (fiducial) & on & 18 au & 0.07 & 0.066 & -2.25 & -2.25 & -0.5 & -0.6\\
        \texttt{54au-rad} & on & 54 au & 0.07 & 0.11 & -2.25 & -2.25 & -0.5 & -0.65\\
        \texttt{18au-lowdens-rad} & on & 18 au & 0.07 & 0.12 & -2.25 & -2.5 & -0.5 & -0.36\\
        \hline
        \texttt{18au-iso} & off & - & 0.066 & - & -2.25 & - & -0.6 & - \\
        \texttt{18au-bkgT} & off & - & - & - & - & - & - & - \\
        \texttt{18au-bkgT-bkgCool} & off & - & - & - & - & - & - & - \\
        \texttt{54au-bkgT-bkgCool} & off & - & - & - & - & - & - & - \\
	\hline
	\end{tabular}
 \label{tab:models}
\end{table*}

The gas surface density profile  follows a power-law with an exponential cutoff, consistent with viscous evolution models \citep{lynden-bell74, hartmann98} and observational constraints \citep[e.g., ][]{miotello22}. The gas surface density is given by:

\begin{equation}
    \Sigma_\mathrm{g} = \Sigma_\mathrm{g,0} \mathrm{(R/R_0)^{-1}}\ \mathrm{exp(-(R-R_0)/100\ au)},
	\label{eq:surface density}
\end{equation}
where $\Sigma_\mathrm{g,0}$ is the gas surface density at a reference radius of $\mathrm{R_0}$ = 1 au. In our fiducial models, $\Sigma_\mathrm{g,0}$ is set to $178\ \mathrm{g\ cm^{-2}}$, following \citet{zhu12}.

Regarding the initial temperature structure, we assumed a radial power-law profile as the initial condition:
\begin{equation}
\label{eq:temperature}
T(R,Z)=T(R_{0})\left(\frac{R}{R_{0}}\right)^q\,,
\end{equation}
where $q$ is set to -0.5. The reference temperature $T(R_{0})$ is given by:
\begin{equation}
    T (R_0) = \Big(\frac{fL_*}{4\pi R_0^2\sigma_{b}} \Big)^{1/4}.
\end{equation}
Here, $f$ accounts for the flaring of the disk \citep[e.g.,][]{chiang97, dalessio98, dullemond01}, and we used a value of $f=0.1$ in our initial conditions. The stellar luminosity $L_*$ is assumed to be 1 $L_\odot$, and $\sigma_{b}$ represents the Stefan-Boltzmann constant.

Then the hydrostatic equilibrium in the $R-Z$ plane requires the initial density profile at the disk midplane to be \citep[e.g.,][]{nelson13}
\begin{align}
\rho_{0}(R,Z=0) \ \ \ \ \ \ \ \ \ \ \ \ \ \ \ \ \ \ \ \ \ \ \ \ \ \ \ \ \ \ \ \ \ \ \ \ \ \ \  \ \ \ \ \ \ \ \ \ \ \ \ \ \nonumber\\
=\rho_{0}(R=R_{0},Z=0)\left(\frac{R}{R_{0}}\right)^p \mathrm{exp\left((R_0-R)/100\ au\right)} \,,\label{eq:rho}
\end{align}
where at the initial condition, the gas surface density and the volume density is related by $\rm \rho_{0}(R,Z=0) = (2\pi)^{-1/2}\Sigma_g(R)$$h$$\rm(R)^{-1}$. The midplane radial density profile power-law index $p$ can be related to the surface density power-law index $r$ ($\Sigma_g \propto R^{r}$) and temperature power-law index $q$, by $p = r - q/2 - 3/2$. Since we adopted $q=-0.5$ and $r=-1$, $p$ = -2.25. The temperature is used to calculate the gas scale height $h$ = $c_s/\Omega_K$, where $c_{s}= (P/\rho)^{1/2}$ is the isothermal sound speed, and $\Omega_K$ = $(GM_{*}/R^3)^{1/2}$ is the Keplarian orbital frequency. We adopt $M_*$ = 1 $M_\odot$.

In the vertical direction
\begin{equation}
\rho_{0}(R,Z)=\rho_{0}(R,Z=0) {\rm exp}\left[\frac{GM}{c_{s}^2}\left(\frac{1}{\sqrt{R^2+Z^2}}-\frac{1}{R}\right)\right]\,,\label{eq:rho0}
\end{equation}
and the azimuthal velocity
\begin{equation}
\vel_{\phi}(R,Z)=\vel_{K}\left[(p+q)\left(\frac{c_{s}}{\vel_{K}}\right)^2+1+q-\frac{qR}{\sqrt{R^2+Z^2}}\right]^{1/2}\,,\label{eq:vphi}
\end{equation}
where $\vel_{K}=\Omega_{K}R=\sqrt{GM_{*}/R}$ \citep[e.g.,][]{nelson13}. We can see that the vertical shear rate $d\vel_{\phi}/dZ$ is non-zero as long as $q\neq 0$.
The other two velocity components $\vel_R$ and $\vel_Z$ are set to be zero at the initial condition.
We do not consider the self-gravity of the disk in this paper, which should be a valid assumption for most of the Class II disks \citep{miotello22}.

In the initial condition (Equation \ref{eq:temperature}), we also assumed the temperature to be vertically isothermal. This assumption is valid near the midplane. However, we will demonstrate that the quasi-steady state of our simulations exhibits a cool midplane and a superheated atmosphere which is consistent with classical analytical calculations \citep{calvet91,chiang97, dalessio98}, Monte Carlo radiative transfer calculations \citep{pascucci04, pinte09}, previous radiation hydrodynamic simulations \citep{flock13, flock17, flock20, kuiper10, kuiper13}, and recent ALMA CO observations \citep{law22, law23}. We will also show that this vertical temperature gradient, together with the varying local orbital cooling time, is crucial for the gas kinematics.

The full disk corresponds to a value of $\rm r_{cav}$ equal to 3 solar radii (\texttt{3$\mathrm{r_\odot}$-rad}), which represents the magnetic spherical accretion truncation radius \citep[e.g., ][]{hartmann16}. The total gas mass is approximately 0.01 solar masses (0.007 solar masses for \texttt{54au-rad}). The transition disks have $\rm r_{cav}$ values of 18 au (\texttt{18au-rad}, fiducial model) and 54 au (\texttt{54au-rad}). Additionally, we considered a case where the gas surface density is reduced to 1\% of the fiducial value, i.e., $\Sigma_\mathrm{g,0}$ is set to $1.78\ \mathrm{g\ cm^{-2}}$ (\texttt{18au-lowdens-rad}). In this low-density disk scenario, the total gas mass is approximately $10^{-4}$ solar masses. For transition disk with r$_\mathrm{cav}$ = 54 au, we used a \texttt{tanh} profile to make a smooth transition at the inner gap so that it satisfies the Rayleigh criterion \citep[e.g.,][]{yang10} at the initial condition. We summarize all our models in Table \ref{tab:models}.

Limited by the computational cost, the inner boundary of the disk cannot be too small. Thus, we set the inner boundary at 21.6 au. For $r_\mathrm{cav}$ = 3 $\mathrm{r_\odot}$, and 18 au disks, the simulation inner boundaries are beyond the cavity sizes. To mimic the optical depth effect of these disks, we artificially added optical depth between 3 $\mathrm{r_\odot}$/18 au and the simulation's inner boundary at 21.6 au (see Equation \ref{eq:TAU} in Appendix \ref{sec:raytracing}). However, this preset optical depth cannot adjust its vertical structure self-consistently and can lead to discontinuities at the simulation's inner boundary. The direct irradiation on disk inner cavity is also related to the shadowing effect in \citet{dullemond01, jang-condell12, jang-condell13, siebenmorgen12, zhang21}. Therefore, what occurs near the inner boundary might not be reliable. 

\subsection{Radiation Hydrodynamics}
We introduce the numerical setup for the radiation hydrodynamic simulations using the frequency-integrated (gray) radiation module \citep{jiang21}. We detail the implementation of the stellar irradiation and unit conversion in Appendix \ref{sec:raytracing}.

We adopted the Courant–Friedrichs–Lewy (CFL) number to be 0.4, and used second order Van Leer time integrator (vl2), second order spatial reconstruction, and HLLC Riemann Solver. We adopted adiabatic index $\gamma_g$ = 1.4. We discretized the radial direction into 1568 cells, logarithmically spaced from 0.54 to 8 times the reference radius (r$_0$ = 40 au, so 21.6 au to 320 au from inner and outer boundaries). The polar direction was divided into 1536 cells, covering a range from 0.383 to 2.76 radians (68$^\circ$ above and below the midplane). For our fiducial model, this amounts to 45 cells per scale height at $r_0$ (40 au). For the hydro boundary conditions, we used outflow at the inner boundary, and copied initial conditions for outer, upper and lower boundaries. As for radiation boundary conditions, light beams can freely transport out of the domain. If the beam points inward the computational domain, the radiation is assumed to have the background temperature (10 K = 1.63$\times 10^{-3}$ $T_0$, where $T_0$ is the temperature in code unit), which is a typical temperature of molecular clouds. We adopted periodic boundary condition in the azimuthal $\phi$-direction.

The radiation transport uses discrete ordinate, where rays are discretized into different angles. We used the discretization better suited for curvilinear coordinates (\texttt{angle\_flag} = 1) and set \texttt{nzeta} = 2, \texttt{npsi} = 2, where \texttt{nzeta} represents angles from 0 to $\pi$/2 in $\zeta$ direction and \texttt{npsi} represents 0 to $\pi$ in $\psi$ direction. Here $\zeta$ and $\psi$ are the polar and longitudinal angles with respect to the local coordinate, so these angles can point to different directions at different spatial locations \citep{jiang21}. There are 16 angles in total. To test the convergence of the temperature calculated by different numbers of rays, we froze the hydrodynamics (assuming a static disk) and tried \texttt{nzeta} = 4, \texttt{npsi}  = 4, and \texttt{nzeta} = 8, \texttt{npsi} = 8. We found convergence when \texttt{nzeta} = 4 and \texttt{npsi} = 4. Since the temperature difference is already small between \texttt{nzeta} = \texttt{npsi} = 2 and \texttt{nzeta} = \texttt{npsi} = 4, we adopted the former to save computational time.

We ran simulations with \texttt{cfl\_rad} = 0.3 (\texttt{cfl\_rad} is an additional factor multiplied in front of the CFL number to help convergence for implicit method), reduced speed of light $\Rrat =$ 4 $\times$ 10$^{-3}$ \citep{zhangd18, zhu20}, and \texttt{error\_limit} = $10^{-3}$ for the first 10 orbits (P$_\mathrm{in}$, the orbital period of the inner boundary at 0.54 $r_0$ or 21.6 au) to approach the quasi-steady state. Otherwise the iteration times or errors were extremely large. Then we restarted the simulation and changed \texttt{cfl\_rad} and $\Rrat$ back to 1. Thus, we do not use reduced speed of light for the longer time evolution for our simulations. We also changed the \texttt{error\_limit} to $10^{-5}$ after the restart. We note that after the restart it only took one iteration to reach the error limit and the typical error was only $10^{-7}$-$10^{-6}$ .

\subsubsection{Dust Opacity Setup}

While the radiation module can treat isotropic scattering properly, we neglected dust scattering and only considered absorption opacity in this paper to better compare with previous isothermal simulations. We also used the frequency-integrated (gray) radiation transport, but we note that multi-group radiation module is also available \citep{jiang22}. Both dust scattering and multi-frequency radiative transfer will be considered in a future publication.

\begin{figure}
\includegraphics[width=\linewidth]{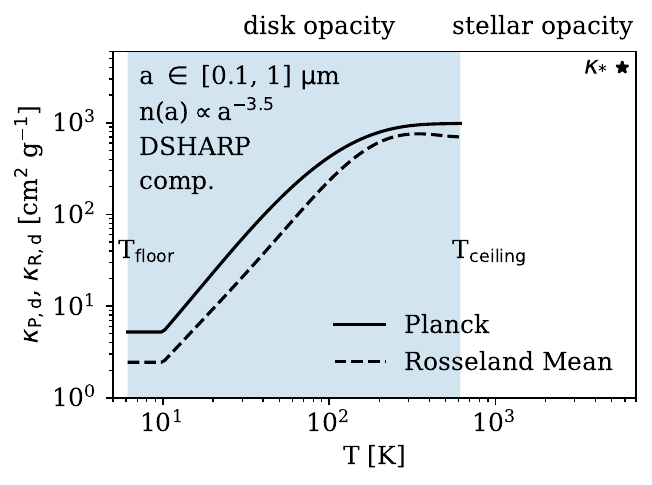}
    \caption{The temperature-dependent dust opacities adopted for all the radiation-hydro models. The solid line indicates the Planck opacity of the disk, whereas the dashed line indicates the Rosseland mean opacity of the disk.  The wavelength-dependent DSHARP opacity \citep{birnstiel18} is convolved at different temperatures to obtain temperature-dependent mean opacities. The stellar temperature is assumed to be 1 T$_\odot$. Its Planck opacity and Rosseland mean opacity are assumed be the same and marked by the star.}
    \label{fig:opac}
\end{figure}

We used the DSHARP composition \citep{birnstiel18} and a power law MRN dust size distribution \citep[$n(a) \propto a^{-3.5}$,][]{mathis77}. The minimum grain size $a_\mathrm{min}$ = 0.1 $\mu$m and maximum grain size $a_\mathrm{max}$ = 1 mm. In our fiducial models, we assumed that only small grains determine the temperature distribution due to their high opacity at the peak of the stellar spectrum; therefore, we considered grains sized between 0.1 and 1 $\mu$m, which account for f$_\mathrm{s}$=0.02184 of the total dust mass. The mass ratio between all the dust and gas was assumed to be 1/100. Then we calculated the Planck and Rosseland mean opacities normalized to the total dust mass ($\kappa_{P,d}$, and $\kappa_{R,d}$) at various disk temperatures and fitted by univariate spline functions labeled as solid and dashed curves, respectively in Figure \ref{fig:opac}. We also calculated the stellar Rosseland mean opacity normalized to the total dust mass ($\kappa_{*,d}$ = 3995 $\mathrm{cm^2 g^{-1}}$) at the solar temperature labeled as the star legend. These opacities can be simply converted to the ones normalized to gas ($\kappa_{P,g}$, $\kappa_{R,g}$, and $\kappa_{*,g}$) by multiplying the dust to gas mass ratio, which is 0.01 for all models. Disk opacities are inputs for the radiation module, whereas the stellar opacity is used in the stellar irradiation as an extra heating source term (see Appendix \ref{sec:raytracing}).

\subsubsection{Comparison with RADMC-3D}
\begin{figure}
\includegraphics[width=\linewidth]{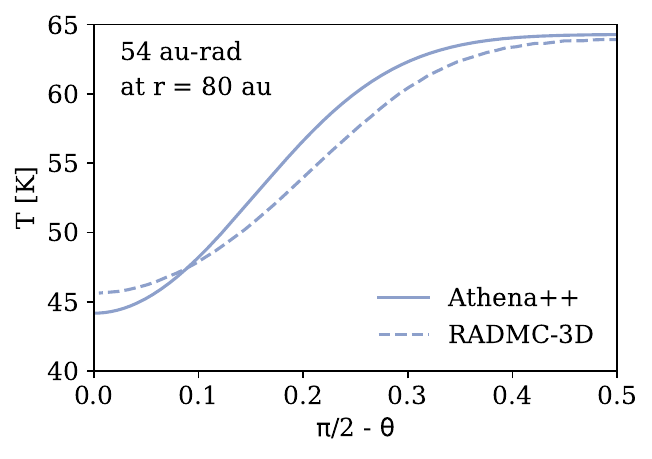}
    \caption{Temperature comparison between Athena++ and RADMC-3D for the transition disk model (\texttt{54au-rad}) at r = 80 au in the vertical direction. The solid and dashed curves show the temperatures in $\theta$ direction for Athena++ and RADMC-3D, respectively.}
    \label{fig:temperature_compare}
\end{figure}

We froze the hydrodynamics and used the initial conditions of full and transition disk models to test the temperature calculation of our irradiation implementation. Then we compared the results with Monte Carlo radiative transfer code RADMC-3D \citep{dullemond12} using the same density structure. In RADMC-3D, we also used the same DSHARP opacity with the same dust properties and dust scattering turned off. When the disk is mostly optically thin to the stellar irradiation, the difference is at most $\sim$7\%, such as in the transition disk with $r_\mathrm{cav}$ = 54 au (shown in Figure \ref{fig:temperature_compare}).

The difference comes from the frequency integrated radiative transfer in Athena++ module and the more accurate multi-frequency treatment in RADMC-3D. This difference has been extensively studied in \citet{kuiper10, kuiper13}, where they found that the temperature calculated by the gray radiation transfer can be underestimated in the optically thick regime and overestimated in the optically thin regime for the stellar irradiation, which is consistent with our results. This is because the region near the midplane is optically thick to the stellar irradiation, so the heating comes from the $\tau_*$ = 1 $(\tau_*$ is the stellar optical depth integrated in the radial direction, see Equation \ref{eq:TAU}) surface in the atmosphere \citep{calvet91,chiang97}. However, even though the stellar spectrum peaks at optical to UV wavelengths, the continuum stellar spectrum still has a set of $\tau_*$ = 1 surfaces for each frequency instead of a single one. The $\tau_*$ = 1 surfaces at longer wavelengths can penetrate deeper and transport more energy to the midplane, thus increasing the temperature at the optically thick region \citep{kuiper10}. They also demonstrated that even within the single frequency radiation-hydro framework, the temperature calculation can be as accurate as the multi-frequency one by integrating the multi-frequency stellar irradiation to mimic continuous $\tau_*$ = 1 surfaces (``hybrid method'' therein). This treatment has also been implemented and tested for our problem and can be used in our future projects.   For the full disk model ($r_\mathrm{cav}$ = 3r$_\odot$) in the current paper, the temperature at the midplane can be underestimated by 40\%, due to the very high optical depth. Therefore, processes that are sensitive to the absolute temperature values (e.g., chemistry) need the hybrid method \citep{kuiper10, kuiper13} or the multi-band radiation module \citep{jiang22} in the optically thick regime.

\subsection{Pure Hydro Simulations with Different Levels of Simplifications}

Since most of the understanding on VSI was from previous vertically isothermal simulations and linear theory, we also ran various pure hydro simulations with different levels of simplifications to compare with our rad-hydro simulations. Namely, they are (a) vertically isothermal simulations with adiabatic EoS and instant cooling ($\beta$ = 10$^{-6}$), (b) vertically varying background temperature with adiabatic EoS and instant cooling, and (c) varying background temperature with adiabatic EoS and local orbital cooling. Their model names are also listed in Table \ref{tab:models}.

\begin{figure}
	\includegraphics[width=\linewidth]{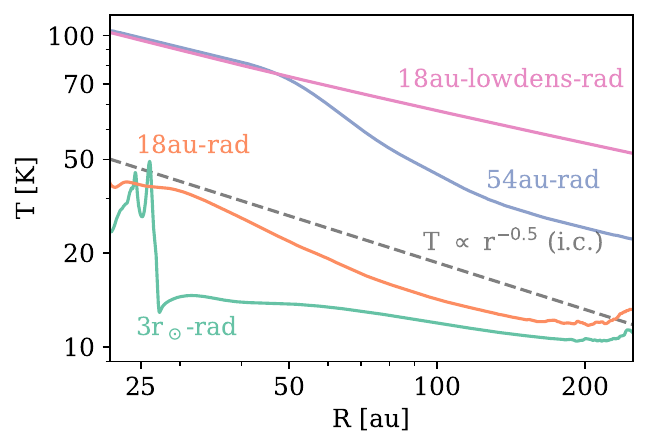}
    \caption{The midplane temperature profiles for four radiation-hydro models (time-averaged from t = 1000-1200 P$_{\mathrm{in}}$, and t = 500-700 P$_{\mathrm{in}}$ for \texttt{54au-rad}). Models for \texttt{3r$_\odot$-rad}, \texttt{18au-rad}, \texttt{54au-rad}, \texttt{18au-lowdens-rad} are shown in green, orange, purple, and magenta lines, respectively. The dashed line indicates the initial condition of the temperature, which is proportional to R$^{-0.5}$.}
    \label{fig:1DradialT}
\end{figure}

\textit{(a) Vertically Isothermal Simulations}.
We tried to compare the vertically isothermal simulations using the same disk aspect ratio ($h/r$) as the rad-hydro simulations, as the Reynolds stress is dependent on $h/r$ \citep{manger20} and also temperature power-law index $q$ \citep{manger21}. Since radiation hydro simulation will adjust the temperature to reach hydrostatic equilibrium from the initial condition, the disk scale height can change from initial condition depending on the radial optical depth of the star. The midplane and atmosphere also have different temperatures. Thus, we measured the midplane temperature and surface density at $r_0$ (40 au), and radial power-law density and temperature indices ($p$ and $q$) of the rad-hydro simulations and put them as initial conditions for these vertically isothermal simulations. The midplane temperatures for four radiation hydro simulations are shown in Figure \ref{fig:1DradialT}, along with their initial condition (dashed line). 

\textit{(b) Background Temperature with Isothermal EoS}.
We used the R-Z two dimensional background temperature averaged between t = 1000-1200 P$_\mathrm{in}$ (500-700 P$_\mathrm{in}$ for $r_\mathrm{cav}$ = 54 au) from rad-hydro simulations and fixed them throughout the simulation. The gas density will adjust according to the temperature profile after the simulation begins.

\textit{(c) Background Temperature with Local Oribital Cooling}.
For these simulations, we used the R-Z background temperature as (b), but used adiabatic EoS with local orbital cooling, where the cooling means that the temperature will be relaxed to the background temperature in a dimensionless cooling time $\beta$ (the cooling time normalized by the Keplerian orbital frequency). We used the simple optically thin and thick cooling times \citep{flock17},
\begin{equation}
    \beta = \beta_\mathrm{thin} + \beta_\mathrm{thick} = \frac{c_v\Omega_K}{16\sigma_b T^3}\big(\kappa_{P,g}^{-1}(T) + 3(h/\hat{k})^2\rho^2\kappa_{R,g}(T)\big),
    \label{eq:beta}
\end{equation}
where $c_v$ = $(\gamma_g-1)^{-1} k_b/\mu m_H$ is the specific heat capacity at constant volume. We estimated the disk scale height by,
\begin{equation}
    h = h_0 (r/r_0)^{1.5+q/2} 
\end{equation}
where $q$ is the midplane temperature power-law slope in the radial direction shown in Table \ref{tab:models}. We assumed $\hat{k}$ = 10 as $h$/10 is a typical length scale measured in \citet{lin15}. However, since the optically thin term ($\beta_\mathrm{thin}$) dominates in most of the region, this adoption of $h$ and $\hat{k}$ is not critical. By combining realistic vertical temperatures and location dependent cooling times, these simulations should have the closest thermodynamical properties compared with rad-hydro simulations, as we will demonstrate in the next section.

\section{Results}
\label{sec:results}

\subsection{Overview}
\label{sec:overview}

\begin{figure*}
	\includegraphics[width=\linewidth]{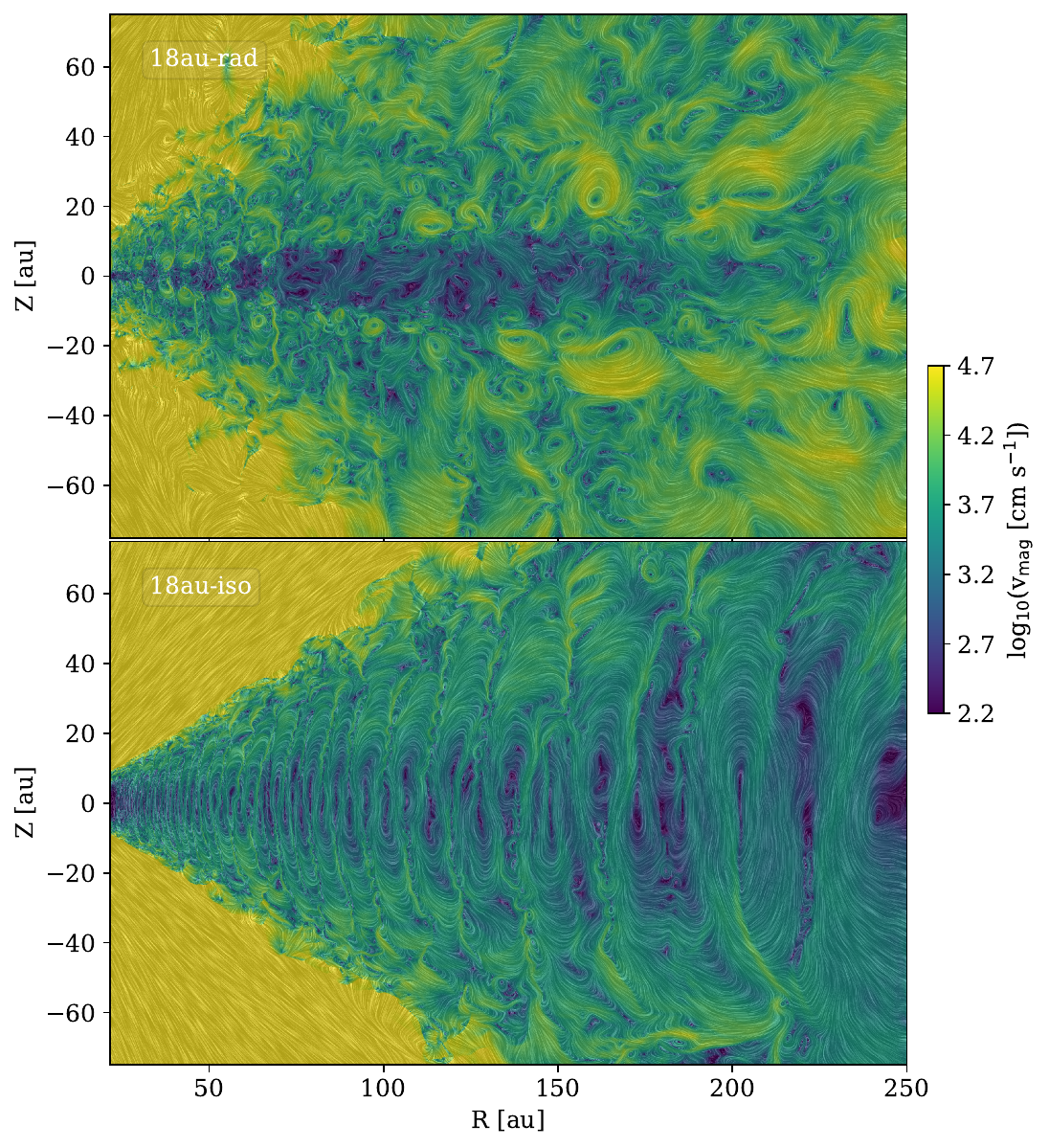}
    \caption{The line integral convolution (LIC) of the velocity field in the meridional plane, (v$_R$, v$_Z$), color-coded by its magnitude, v$_\mathrm{mag}$ for fiducial ($\mathrm{r_{cav}}$ = 18 au) radiation-hydro (top, \texttt{18au-rad}) and vertically isothermal (bottom, \texttt{18au-iso}) models. The flow pattern in the bottom panel is very similar to Figure 7 in \citet{flores-rivera20}.}
    \label{fig:moneyplot}
\end{figure*}

The significant difference between the rad-hydro simulation (top panel) and classical vertically isothermal simulation (bottom panel) can be demonstrated in Figure \ref{fig:moneyplot}, where we show the line integral convolution\footnote{The line integral convolution is a texture-based technique to visualize the vector field without the need to set start and end points, in contrast to a streamline plot. We used python package \texttt{lic} (\url{https://gitlab.com/szs/lic}).} \citep[LIC, similar to][]{flores-rivera20} of our fiducial models ($\mathrm{r_{cav}}$ = 18 au), $\texttt{18au-rad}$ and $\texttt{18au-iso}$ at t = 1000 P$_\mathrm{in}$, color-coded by the meridional velocity, $\big($v$_R^{2}$+v$_Z^{2}$$\big)$$^{1/2}$. In the vertically and locally isothermal simulation (bottom panel), the classical corrugation mode (the radially narrow, vertically extended circulation pattern) is clear. In the rad-hydro simulation (top panel), the disk is separated in two parts, the cool midplane and the superheated atmosphere. In the cool midplane, the velocity and turbulence levels are low, whereas the superheated atmosphere is more turbulent. The boundary between the cool midplane and superheated atmosphere exists a strong shear that leads to many small-scale vortices. The global circulation pattern that can be easily identified in locally and vertically isothermal simulation is replaced by turbulence on smaller scales.

Next, we analyze all of our radiation models in detail accompanied by pure-hydro simulations with various levels of simplifications. These quantities are taken either at $t$ = 1000 $\mathrm{P_{in}}$ or time-averaged values between 1000-1200 $\mathrm{P_{in}}$. In the case of the $\mathrm{r_{cav}}$ = 54 au transition disk, the flow at the cavity edge will reach the critical condition for the Rayleigh stability criterion and become unstable. A giant vortex develops at $\sim$ 800 orbits, which should break into smaller vortices in realistic 3D disks. Therefore, we analyze this particular model from 500 to 700 P$_{\mathrm{in}}$ (t=500 P$_{\mathrm{in}}$ for the snapshot) to avoid this unphysical feature in 2D.

\subsection{Thermal Structure Determines Kinematics}
We will use four rad-hydro simulations to demonstrate that $\tau_*$ = 1 surface ($\tau_*$: radial stellar optical depth, see Equation \ref{eq:TAU}) sets disk temperature and equivalent local orbital cooling structures. Subsequently, these two structures determine the disk kinematics. 

\begin{figure*}
	\includegraphics[width=\linewidth]{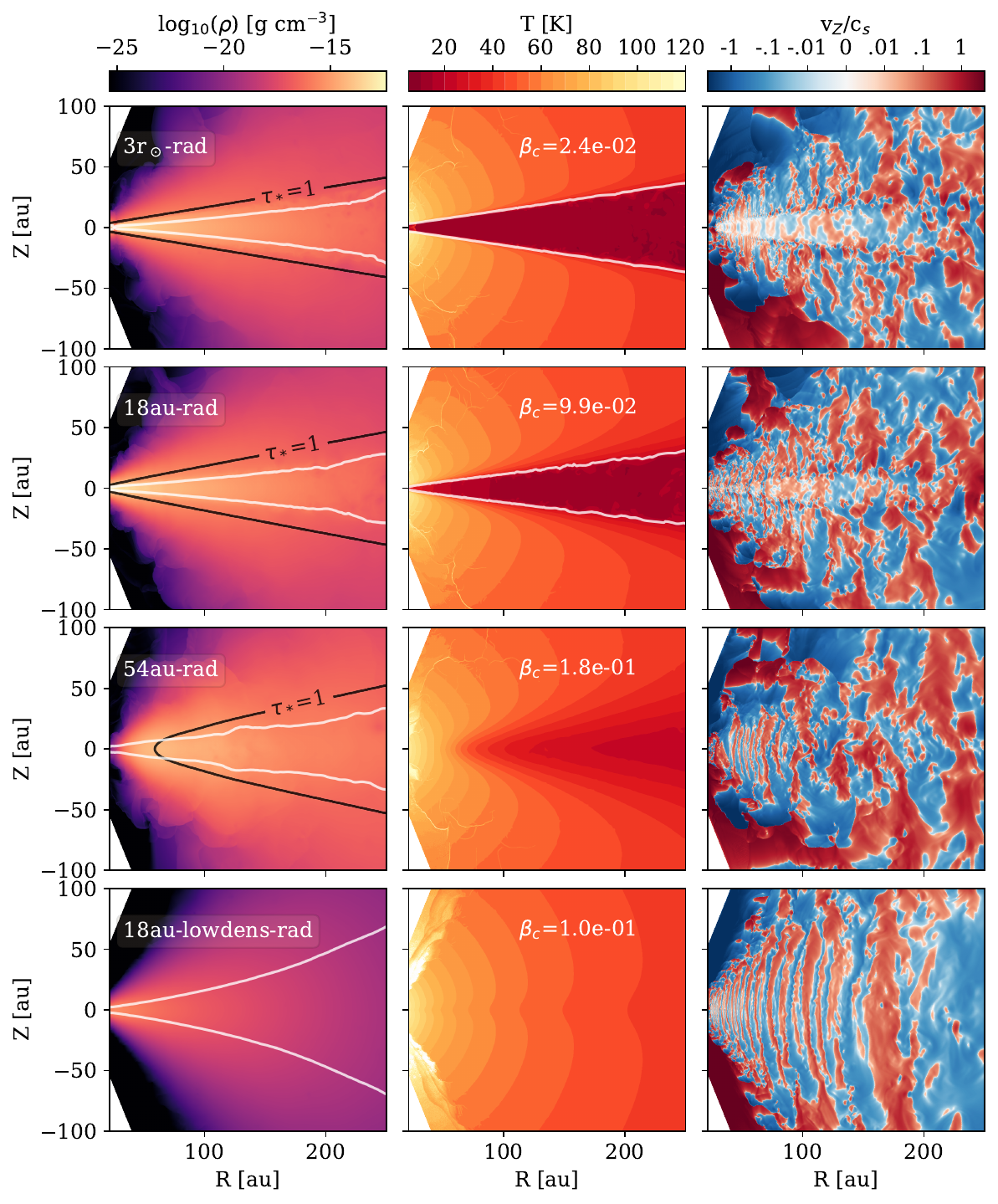}
 \vspace{-2em}
    \caption{From left to right: the gas density ($\rho$), temperature (T), and vertical velocity ($v_Z$) of four radiation-hydro models at t = 1000 P$_{\mathrm{in}}$ (at t =500 P$_{\mathrm{in}}$ for r$_\mathrm{cav}$ = 54 au, \texttt{54au-rad}) in the meridional (R-Z) plane. From top to bottom: the full disk model with r$_\mathrm{cav}$ = 3 r$_\odot$ (\texttt{3r$_\odot$-rad}), the transition disks with r$_\mathrm{cav}$ = 18 au (\texttt{18au-rad}), r$_\mathrm{cav}$ = 54 au (\texttt{54au-rad}), and 1\% of the fiducial density (\texttt{18au-lowdens-rad}). The white contours on the density maps mark the one gas scale height. The black contours are the locations where the stellar optical depth in the radial direction reaches unity (the last model, \texttt{18au-lowdens-rad}, has $\tau_* <$ 1 for the whole disk). The $\beta_c$ on the temperature maps is the critical cooling time for VSI, represented by the white contours. The region enclosed by the contour near the midplane has $\beta > \beta_c$, so the VSI should not be operating according to linear analysis. The last two models, \texttt{54au-rad} and \texttt{18au-lowdens-rad}, have $\beta < \beta_c$ for the whole domain.}
    \label{fig:2DrhoTvZ}
\end{figure*}

The 2D (R-Z) snapshots of gas density (left panels), temperature (middle panels), and the vertical velocity (right panels) for four models are shown in Figure \ref{fig:2DrhoTvZ}. The gas density does not deviate significantly from the initial conditions. Gas scale heights are represented by the white contours overlaid on gas densities, calculated by assuming that vertical density follows a Gaussian profile (see Equation \ref{eq:rho0}) and that the temperature is taken at the midplane. In cases where $\mathrm{r_{cav}}$ = 3$\mathrm{r_{\odot}}$ and 18 au (\texttt{3r$_\odot$-rad} and \texttt{18au-rad}), the midplane temperature is lower than the initial condition, whereas the atmosphere temperature is higher, resulting in smaller effective gas scale heights in the midplane (see Table \ref{tab:models} and Figure \ref{fig:1DradialT}) and larger effective gas scale heights in the atmosphere. In contrast, for $\mathrm{r_{cav}}$ = 54 au and 18 au, low-density cases (\texttt{54au-rad} and \texttt{18au-lowdens-rad}), temperatures are higher than the initial condition, leading to larger effective gas scale heights (see Table \ref{tab:models}). The black contours represent $\tau_*$ = 1 surfaces where stellar irradiation intercepts the disk, setting the two temperature structure of the disk (cool midplane and superheated atmosphere). Note that the low density model (\texttt{18au-lowdens-rad}) lacks this surface, meaning that the entire disk is optically thin to stellar irradiation. 

The locations of the $\tau_*$ = 1 surfaces are shown  in the temperature panels (middle panels) in Figure \ref{fig:2DrhoTvZ}. We observe sharp decreases in temperature below these surfaces for the first three models. The low-density model (\texttt{18au-lowdens-rad}) is nearly vertically isothermal because the entire disk is optically thin to stellar irradiation. The white contours on top of the temperature maps indicate where the cooling time, estimated using Equation \ref{eq:beta}, equals the critical cooling time from Equation \ref{eq:beta_c}, denoted as $\beta_c$ in each panel. The radial temperature gradient $q$ is estimated by fitting a power-law to the midplane temperature profile (see Table \ref{tab:models} and Figure \ref{fig:1DradialT}). The $\mathrm{r_{cav}}$ = 3 $\mathrm{r_\odot}$ and 18 au models exhibit cooling times exceeding the critical cooling time in the cool midplane indicating stability, while the superheated atmosphere has cooling times less the critical cooling time, indicating that the region is unstable to the VSI. The $\mathrm{r_{cav}}$ = 54 au and low-density models do not have cooling times exceeding $\beta_c$, indicating instability throughout the domain.

The vertical velocity ($v_Z/c_s$) structure in the right panels of Figure \ref{fig:2DrhoTvZ} also reflects temperature and cooling time structures. For the $\mathrm{r_{cav}}$ = 3$\mathrm{r_\odot}$ and 18 au models, the vertical velocity is below 1\% of the local sound speed in the cool midplane and above 10\% of the local sound speed in the superheated atmosphere. The separation occurs around a gas scale height and slightly below the $\tau_*$ = 1 surface. In the atmosphere, the vertical velocity is still vertically extended and radially narrow, similar to the n=1 corrugation mode, but the upper and lower disk vertical velocities tend to have opposite signs, differing from the classical corrugation mode. We show in Appendix \ref{sec:anticorrelation} (Figure \ref{fig:cross_correlation}) that these velocities tend to be anti-correlated. The classical n=1 corrugation mode only dominates when the disk has a vertically constant temperature and a short cooling time ($\beta < \beta_c$), which is the case inside the inner cavity of the $\mathrm{r_{cav}}$ = 54 au model and throughout the low-density model.

For the full disk model ($\mathrm{r_{cav}}$ = 3 $\mathrm{r_\odot}$), we observe strong density and velocity perturbations near the inner boundary. This is a simulation artifact because our simulation domain cannot extend to 3 $\mathrm{r_\odot}$ due to the computational cost associated with the very large dynamical range (see the end of Section \ref{sec:disksetup}).

\begin{figure*}
	\includegraphics[width=\linewidth]{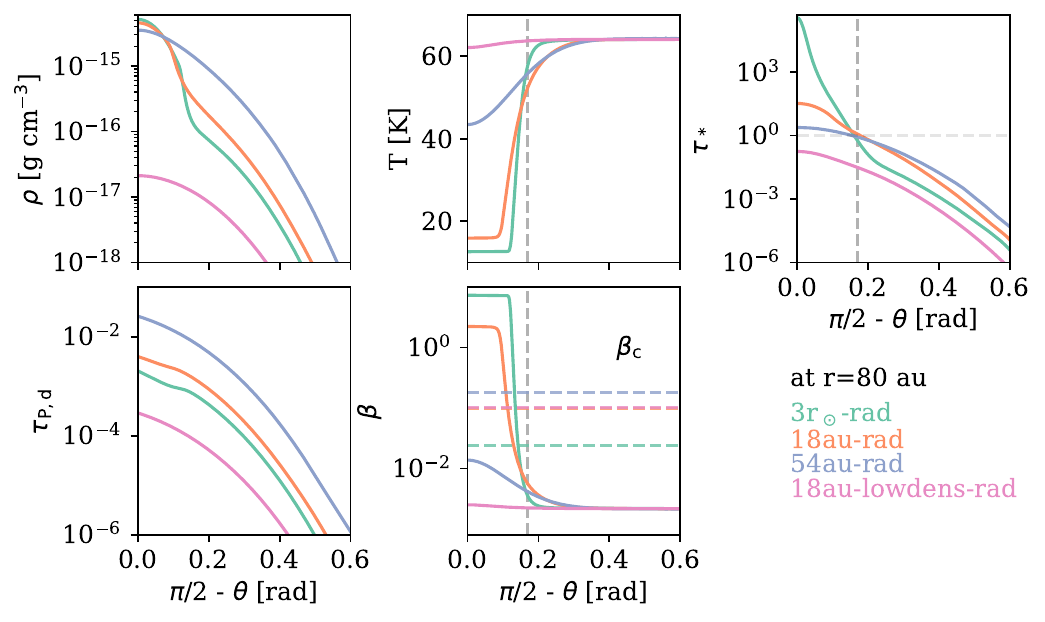}
    \caption{Time-averaged (t = 1000-1200 P$_{\mathrm{in}}$ and 500-700 P$_{\mathrm{in}}$ for $\mathrm{r_{cav}}$ = 54 au, \texttt{54au-rad} model) gas density, temperature, stellar optical depth in the radial direction ($\tau_*$), disk optical depth in the vertical direction ($\tau_{\mathrm{P,d}}$), and the cooling time ($\beta_c$) for four radiation-hydro models cut at r = 80 au in the vertical direction. Models for \texttt{3r$_\odot$-rad}, \texttt{18au-rad}, \texttt{54au-rad}, \texttt{18au-lowdens-rad} are shown in green, orange, purple, and magenta lines, respectively. The horizontal dashed lines in the last panel indicate the critical cooling time, $\beta_{\mathrm{c}}$. The vertical dashed lines indicate the $\tau_*$ = 1 surface for the first three models.}
    \label{fig:1DverticalProfiles1}
\end{figure*}

To quantify the vertical structure of the disk in these radiative-hydrodynamic models, we present time-averaged vertical profiles of gas density, temperature, the radially integrated stellar optical depth ($\tau_*$), vertically integrated disk optical depth ($\tau_\mathrm{P,d}$), and cooling time ($\beta$) at $r = 80$ au in Figure \ref{fig:1DverticalProfiles1}. The densities of the $\mathrm{r_{cav}} = 3\ \mathrm{r_\odot}$ and 18 au models (\texttt{3r$_\odot$-rad} and \texttt{18au-rad}) exhibit two Gaussian distributions, one concentrated at the midplane and the other more extended in the atmosphere. These correspond to the cool midplane and the warmer atmosphere, as shown in the temperature profiles. The temperatures remain almost constant in these two regions, with the transition occurring between 0.1 to 0.3 radians, roughly equivalent to 2-4 gas scale heights. For the $\mathrm{r_{cav}} = 54$ au and low-density models (\texttt{54au-rad} and \texttt{18au-lowdens-rad}), two-Gaussian profiles are not as clear, given their smoother temperature transition. The low-density model is nearly vertically isothermal, with a slight temperature drop at the midplane. The transition between optically thin and thick stellar irradiation occurs at approximately 0.17 radians for all three models at 80 au (indicated by the vertical dashed lines). The low-density case is optically thin to stellar irradiation. This model only contains 10$^{-4}$ M$_\odot$, which falls at the lower end of protoplanetary disk masses. Another way to achieve a very optically thin disk is to modify our assumptions regarding the fiducial small grain fraction and the dust-to-gas mass ratio. If the disk is entirely depleted of small particles or has an extremely low dust-to-gas mass ratio, it can become optically thin to stellar irradiation. The $\tau_{\mathrm{P,d}}$ panel indicates that all models are optically thin to the disk's emission. The dimensionless cooling times for these models range from 10$^{-3}$ to 10. Their critical cooling times, denoted by the horizontal dashed lines, differ due to variations in gas scale heights and radial temperature gradients. For the $\mathrm{r_{cav}} = 3\ \mathrm{r_\odot}$ and 18 au models, the transition between the VSI unstable atmosphere and the VSI stable midplane occurs around 0.15 radians, approximately 2-3 gas scale heights. The $\mathrm{r_{cav}} = 54$ au and low-density models have cooling times smaller than the critical cooling times throughout their vertical extent, suggesting that the VSI should operate along the entire vertical extent.

\begin{figure*}
	\includegraphics[width=\linewidth]{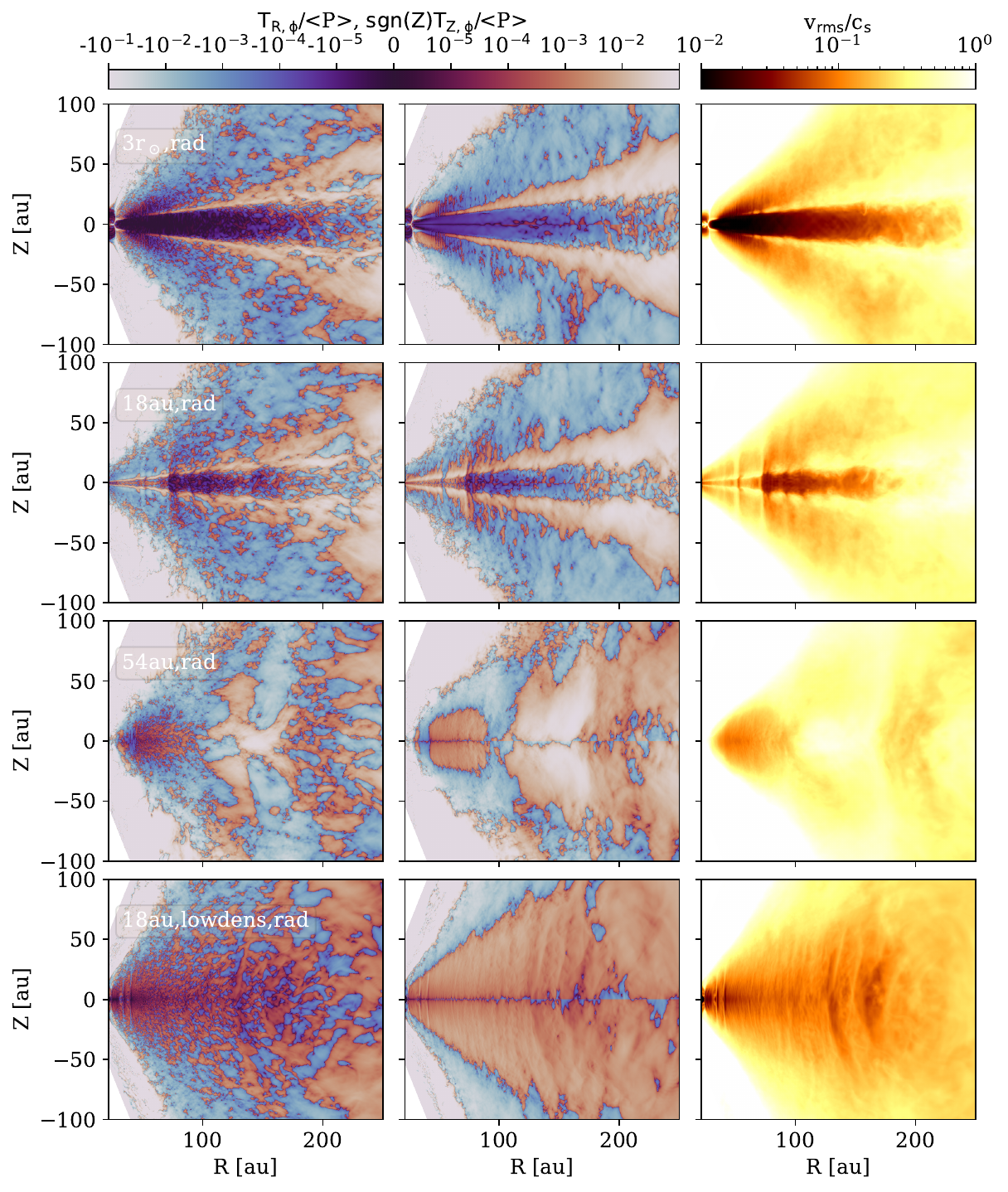}
 \vspace{-2em}
    \caption{From left to right: the turbulence stresses for Z-$\phi$ and  R-$\phi$ components normalized by the averaged gas pressure, and the root mean square velocity normalized by the local sound speed. The values are calculated between t = 1000-1200 P$_{\mathrm{in}}$, (at t =500-700 P$_{\mathrm{in}}$ for r$_\mathrm{cav}$ = 54 au, \texttt{54au-rad}). Other layouts are similar to Figure \ref{fig:2DrhoTvZ}.}
    \label{fig:2Dshear_vrms}
\end{figure*}

To indicate locations where VSI is active or inactive in the R-Z plane, we present R-$\phi$ ($T_\mathrm{R,\phi}$, left panels) and Z-$\phi$ ($T_\mathrm{Z,\phi}$, middle panels) Reynolds stresses, normalized by the time-averaged pressure, along with root mean square velocity normalized by the local sound speed (right panels) for four radiation models in Figure \ref{fig:2Dshear_vrms}.
We define $\alpha_R \equiv T_\mathrm{R, \phi}/\langle P \rangle_\mathrm{t}$ and $\alpha_Z \equiv T_\mathrm{Z, \phi} /\langle P \rangle_\mathrm{t}$, where $T_\mathrm{R, \phi} \equiv$  $\mathrm{\langle\rho v_R v_\phi\rangle_t - \langle v_\phi\rangle_t\langle\rho v_R\rangle_t}$, and $T_\mathrm{Z, \phi} \equiv$  $\mathrm{\langle\rho v_Z v_\phi\rangle_t - \langle v_\phi\rangle_t\langle\rho v_Z\rangle_t}$ \citep[][]{nelson13, stoll14, flock17}. The values are calculated between t = 1000-1200 P$_{\mathrm{in}}$, except for the \texttt{54au-rad} simulation which was taken between t = 500-700 P$_{\mathrm{in}}$. In all three columns, brighter colors indicate higher turbulence values. In the two stress columns, red colors denote positive values, while blue colors indicate negative values. The \texttt{3r$_\odot$-rad} model shows low levels of $\alpha_R$ ($\lesssim 10^{-5}$) and $\alpha_Z$ ($\lesssim 10^{-4}$) in the midplane, but they increase from the inner to the outer disk. The stress becomes much larger around the $\tau_* = 1$ surface and approaches $10^{-2}$ before reversing sign to negative values. Overall, $\alpha_Z$ is larger than $\alpha_R$ near the midplane. The root mean square velocity is at the percent level of the sound speed near the midplane, approaching a fraction of the sound speed in the atmosphere, with the strongest values at the transition region (see Section \ref{sec:accretion_shear}). The \texttt{18au-rad} model exhibits similar behavior. Its midplane turbulence becomes slightly higher and the low turbulence region is more confined to the midplane. The \texttt{54au-rad} model shows VSI-like anisotropic turbulence between $\alpha_R$ and $\alpha_Z$ (similar to \texttt{18au-lowdens-rad}) in the cavity and at the ring until $\sim$ 90 au, where $\alpha_R$ is $\sim 10^{-4}$ and $\alpha_Z$ is $\sim 10^{-2}$. At the outer disk (100-160 au), turbulence levels become much higher ($\gtrsim 10^{-2}$) in both stress components, which are much stronger than the midplane regions in all other models. This suggests that the gap edge in transition disks can be highly turbulent due to direct stellar irradiation. Beyond 160 au, however, the \texttt{18au-rad} model has higher turbulence levels in the midplane, since the more turbulent atmosphere has enough energy to disturb the midplane due to lower density contrast at the outer disk (also see Figure \ref{fig:moneyplot}). For the \texttt{18au-lowdens-rad} model, the turbulence structures are very similar to those of vertically and locally isothermal VSI simulations. The $\alpha_R$ is $\sim 10^{-4}$ and $\alpha_Z$ is $\sim 10^{-2}$ throughout the disk, while the root mean square velocity is at the percent level of the local sound speed globally, yet still higher than the midplane values for \texttt{3r$_\odot$-rad} and \texttt{18au-rad} models.

\subsection{Accretion, Zonal Flow, and Vertical Shear Rate}
\label{sec:accretion_shear}
The stress structure determines the disk's accretion structure. We can reveal this relation by averaging the angular momentum equation in the azimuthal direction \citep[e.g., ][]{turner14, lesur21,rabago21,zhu23} and obtain
\begin{align}
 R\frac{\partial\langle\rho \delta v_{\phi}\rangle}{\partial t} = -\langle\rho  v_{R}\rangle \frac{\partial}{\partial R}\big(R\langle v_{\phi}\rangle\big) - \langle\rho  v_{Z}\rangle R \frac{\partial \langle v_{\phi}\rangle}{\partial Z}\nonumber \\
 -\frac{1}{R}\frac{\partial}{\partial R}\Big(R^2T_{R,\phi}\Big) - R\frac{\partial}{\partial Z}T_{Z,\phi} \ ,
\label{eq:amephi}
\end{align}
where $\delta v_{\phi}$ = $v_{\phi}$ - $\langle v_{\phi}\rangle$, $T_\mathrm{R, \phi} =$  $\mathrm{\langle\rho v_R v_\phi\rangle - \langle v_\phi\rangle\langle\rho v_R\rangle}$, and $T_\mathrm{Z, \phi} =$  $\mathrm{\langle\rho v_Z v_\phi\rangle - \langle v_\phi\rangle\langle\rho v_Z\rangle}$. Since our simulations are in 2D, we calculate time averaged instead of azimuthally averaged quantities. If we adopt a smooth disk structure (Equation \ref{eq:vphi}), the change of $\langle v_\phi \rangle$ is small in $Z$ direction, so the second term in the first line can be neglected. We note that in our rad hydro models, the shear at the transition region between the atmosphere and midplane can be large, but since $\langle \rho v_Z \rangle$ is still small compared to $\langle \rho v_R \rangle$, the second term is still less than the first term. If we also assume the disk reaches a steady state (left hand side is zero), the accretion structure is only determined by the derivatives of the stresses. That is\footnote{Besides angular momentum, we also briefly discuss the energy budget in Figure \ref{fig:energy_budget} in the Appendix.},

\begin{align}
 \langle\rho  v_{R}\rangle \frac{\partial}{\partial R}\big(R\langle v_{\phi}\rangle\big) = -\frac{1}{R}\frac{\partial}{\partial R}\Big(R^2T_{R,\phi}\Big) - R\frac{\partial}{\partial Z}T_{Z,\phi} \ .
\label{eq:amephi2}
\end{align}

\begin{figure*}
	\includegraphics[width=\linewidth]{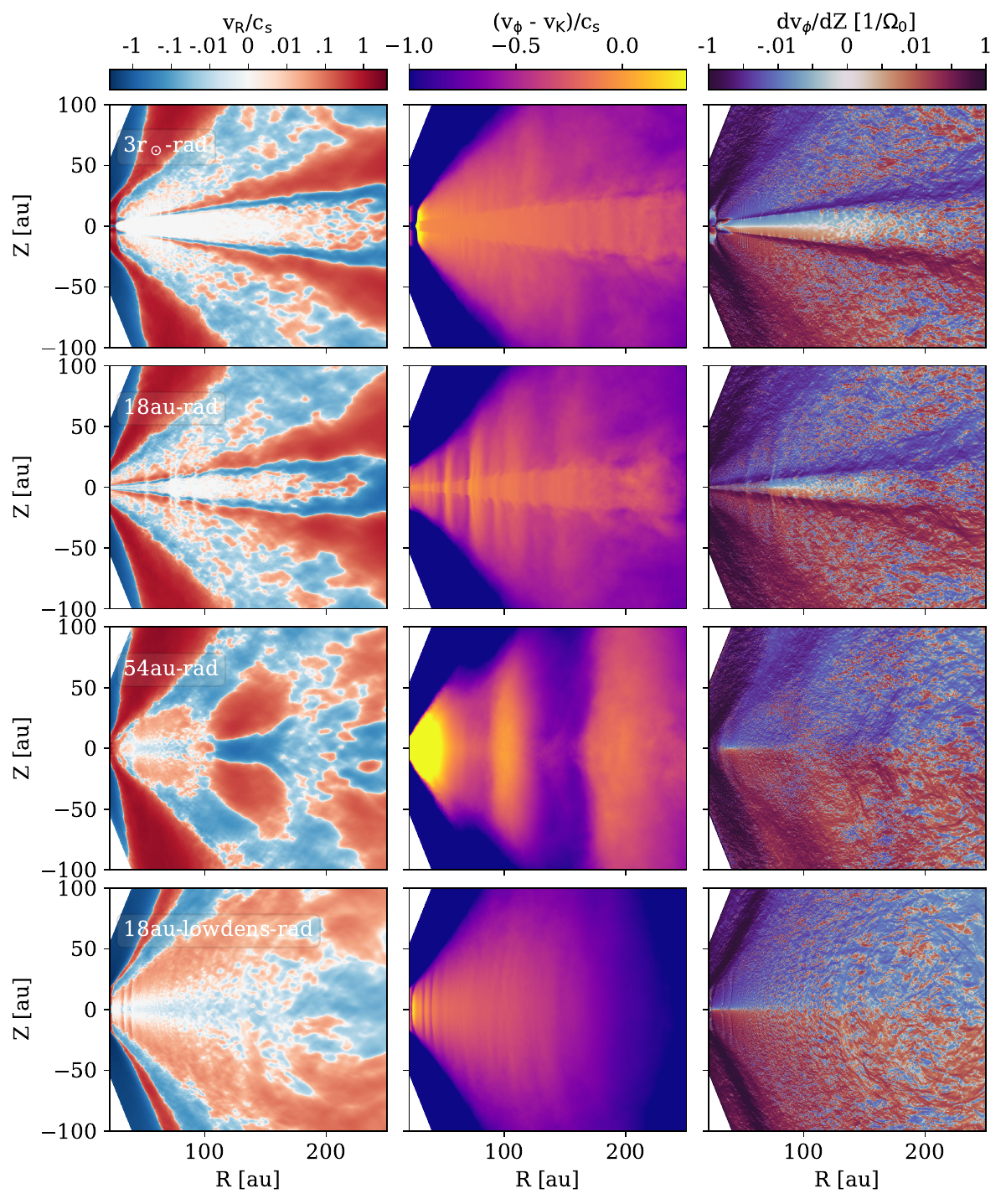}
    \caption{From left to right: the time-averaged (t = 1000-1200 P$_{\mathrm{in}}$ and 500-700 P$_{\mathrm{in}}$ for $\mathrm{r_{cav}}$ = 54 au, \texttt{54au-rad} model) values of the radial velocity (v$_R$), azimuthal velocity subtracted by Keplerian velocity (v$_\mathrm{\phi}$ - v$_\mathrm{K}$), and the vertical shear rate dv$_\mathrm{\phi}$/dZ for four radiation-hydro models in the meridional (R-Z) plane in the same layout as Figure \ref{fig:2DrhoTvZ}.}
    \label{fig:2DvRvphidphidZ}
\end{figure*}

Then we can use Figure \ref{fig:2Dshear_vrms} and Equation \ref{eq:amephi2} to explain the time-averaged radial velocity ($\mathrm{v_R/c_s}$) in Figure \ref{fig:2DvRvphidphidZ} shown on the left panels. Since in all our models the vertical gradient of the stress is greater than the radial gradient among the transition region between midplane and atmosphere, and $T_{Z,\phi}$ is greater than $T_{R,\phi}$, we can just use the second term on the right hand side of Equation \ref{eq:amephi2} and the center column of Figure \ref{fig:2Dshear_vrms} to understand the accretion structure. The radial velocity in the midplane is very small for both $\mathrm{r_{cav}} = 3\ \mathrm{r_\odot}$ and 18 au models due to the small stress and stress gradient there. At the boundary between the cool midplane and superheated atmosphere, an outgoing flow is on top of an ingoing flow, resembling layered accretion. In Figure \ref{fig:2Dshear_vrms}, the region of the ingoing flow is aligned with the sharp transition of negative stress to positive stress above the midplane, whereas the region of the outgoing flow is aligned with the transition of positive stress to negative stress in the upper atmosphere. The sign and magnitude of $\langle \rho v_R \rangle$ can be perfectly calculated from Equation \ref{eq:amephi2}. These regions also align with the regions that have the strongest shear (right panels). In contrast, the midplane shear rate is at least an order of magnitude smaller. The $\mathrm{r_{cav}} = 54$ au and low-density models have ingoing flow in the midplane, and outgoing flow in the atmosphere, which is consistent with previous analytical studies on VSI operating isothermal disks \citep{stoll17a, rabago21}. Such an accretion structure is expected when the turbulence levels between $R-\phi$ and $Z-\phi$ directions are anisotropic, with the former smaller than the latter by around two orders of magnitude. These vertical shear rates for these two models are more uniformly distributed in the vertical direction, due to their more vertically uniform stress profiles which result from smoother temperature profiles. Compared to $\mathrm{r_{cav}} = 3\ \mathrm{r_\odot}$ and 18 au models, the shear rates in these two models are larger in the midplane.

The direction of the radial velocity and shear rate can be intuitively understood by examining the azimuthal velocity shown in the middle panels of Figure \ref{fig:2DvRvphidphidZ} (azimuthal velocity subtracted by the local Keplerian velocity). For the first two models that have temperature stratification, the midplane has faster rotational velocity than the atmosphere. At the transition region near the midplane, the gas loses its angular momentum due to the shear, thus moving inward, whereas the gas near the atmosphere gains its angular momentum and moves outward. In the bottom two models, which lack a strong vertical temperature structure, the azimuthal velocity has less vertical dependence, resulting in significantly smaller shear and radial velocity. Additionally, all four models exhibit zonal flows with perturbations to the azimuthal velocity. The $\mathrm{r_{cav}} = 3\ \mathrm{r_\odot}$ and 18 au models have relatively smaller scale perturbations on several au, whereas the 54 au model has zonal flows on tens of au. The low density model's azimuthal velocity perturbation is small and follows the corrugation pattern for vertical velocity. These zonal flows are associated with substructures in gas (see Section \ref{sec:substructure}) and can be observed in near-infrared scattered light (see Section \ref{sec:observation}).

\begin{figure*}
	\includegraphics[width=\linewidth]{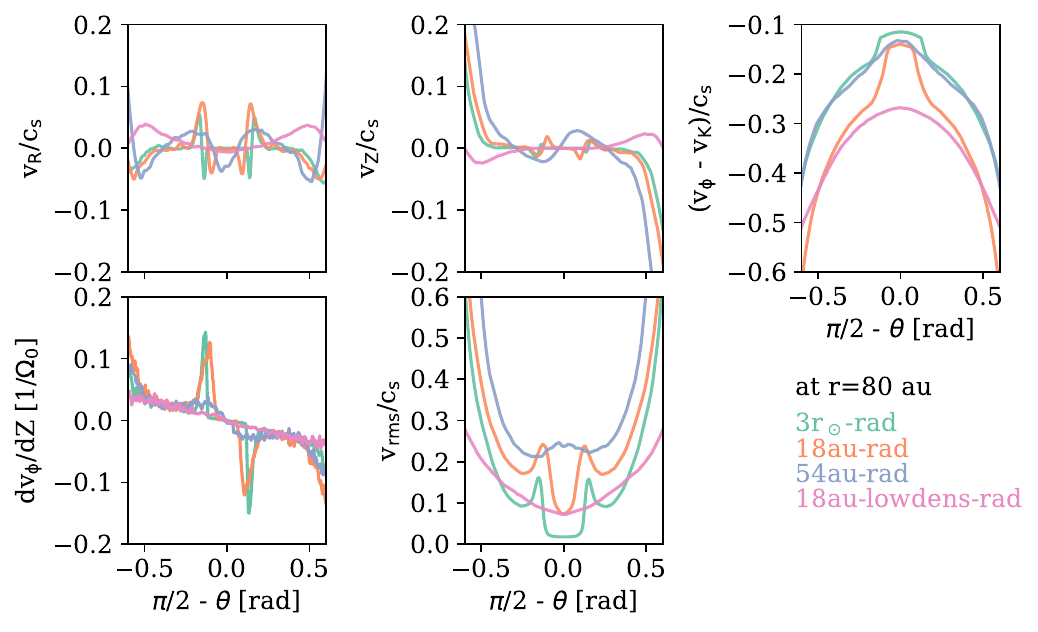}
    \caption{The vertical slices of the time-averaged (from 1000-1200 P$_{\mathrm{in}}$ and 500-700 P$_{\mathrm{in}}$ for $\mathrm{r_{cav}}$ = 54 au, \texttt{54au-rad} model) radial velocity (v$_\mathrm{R}$), vertical velocity (v$_\mathrm{Z}$), azimuthal velocity subtracted by Keplerian velocity (v$_\mathrm{\phi}$ - v$_\mathrm{K}$), vertical shear rate (dv$_\mathrm{\phi}$/dZ), and the root mean square velocity (v$_\mathrm{rms}$) of four radiation-hydro models at r = 80 au.}
    \label{fig:1Dv}
\end{figure*}

To be more quantitative, we present vertical profiles of various time-averaged velocity fields at $r = 80$ au in Figure \ref{fig:1Dv}. Regarding radial velocity, the outgoing and ingoing flows for the $\mathrm{r_{cav}} = 3\ \mathrm{r_\odot}$ and 18 au models are approximately 5\% of the local sound speed, with the outgoing flow having a higher magnitude than the ingoing flow. The $\mathrm{r_{cav}} =$ 54 au model shows ingoing flow in the midplane and outgoing flow above the midplane, both at around 2\% of the sound speed. In contrast, the low-density model displays ingoing flow in the midplane at less than 1\% of the sound speed, and outgoing flow above the midplane, which gradually increases to 4\% of the sound speed at 0.5 radians. The time-averaged vertical velocities in the first three models near the midplane are at the level of 1-2\% of the sound speed, while the low-density model has negligible velocity. As for azimuthal velocity, all models show sub-Keplerian motion due to the radial pressure gradient. The $\mathrm{r_{cav}} = 3 \mathrm{r_\odot}$ and 18 au models exhibit a sharp transition in azimuthal velocity around 0.2 radians, attributed to the steep vertical temperature gradient between the cool midplane and the superheated atmosphere. In the same region, there is a smooth transition in azimuthal velocity for the $\mathrm{r_{cav}} = 54$ au model. The shear rate for the $\mathrm{r_{cav}} = 3\ \mathrm{r_\odot}$ and 18 au models peaks at 0.2 radians, reaching 0.15 $\Omega_{K,0}^{-1}$, where $\Omega_{K,0}^{-1}$ is the time unit, or the inverse of Keplerian orbital frequency at unit radius r$_0$ = 40 au. The $\mathrm{r_{cav}} = 54$ au model also exhibits a slight increase in the same region. In contrast, the low-density model displays a linear increase in vertical shear above and below the midplane. The root mean square of the full velocity is also highest around 0.2 radians for the two models. At the midplane, the $\mathrm{r_{cav}} = 3\ \mathrm{r_\odot}$ model is relatively quiet, with velocities at approximately 2\% of the sound speed. The 18 au model and the low-density model exhibit similar values at the midplane, approximately 10\% of the sound speed. In contrast, the 54 au model is much more turbulent, with velocities exceeding 20\% of the sound speed even at the midplane.

\begin{figure}
	\includegraphics[width=\linewidth]{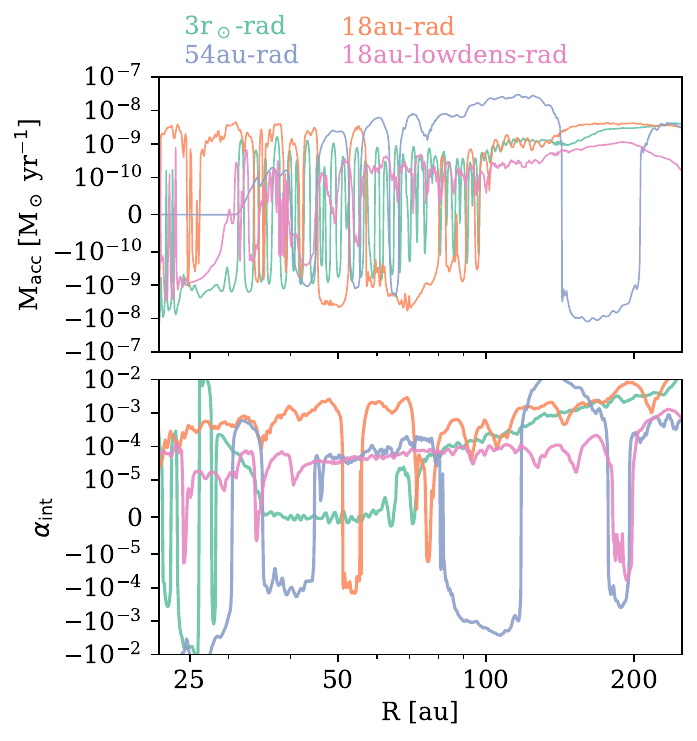}
    \caption{The time-averaged accretion rate in the radial direction (integrated in the vertical direction), and $\alpha_{int}$ (vertically integrated $\alpha_R$) for four radiation-hydro models. The M$_\odot$ yr$^{-1}$ for the low density model is multiplied by 100 to show its value. The color representations are the same as previous figures.}
    \label{fig:accretion_rate_R}
\end{figure}

We also integrated $\langle \rho v_R \rangle$ along the vertical direction to obtain (time-averaged and vertically integrated) radial mass accretion rates
($\dot{M}_{acc}= 2\pi R\int \langle\rho v_{R} \rangle dZ$) as functions of $R$ and show them in Figure \ref{fig:accretion_rate_R} (upper panel). We also show
the vertically integrated $\alpha_{R}$ parameter in the lower panel, defined as
\begin{equation}
\alpha_{int}=\frac{\int T_{R,\phi}dZ}{\int \langle P \rangle dZ}\,.\label{eq:alphaint}
\end{equation}
$\dot{M}_{acc}$ is associated with the radial gradient of $\alpha_{int}$, since if we integrate Equation \ref{eq:amephi2} along $Z$ and assume that $\langle v_\phi \rangle$ equals to the midplane Keplerian speed $v_K$ and does not change with $Z$, Equation \ref{eq:amephi2} can be written as
\begin{align}
&\dot{M}_{acc}=-\frac{2\pi}{\partial R v_{K}/\partial R}\times\nonumber\\
&\left(\frac{\partial}{\partial R}\left( R^2 \alpha_{int}\int \langle P \rangle dZ\right)+R^2T_{Z,\phi}\bigg |_{Z_{min}}^{Z_{max}}\right)\,,\label{eq:mdot}
\end{align}
where the T$_{Z,\phi}$ is typically small at the boundaries due to the small densities at the disk surface. 

The fluctuation is strong in the radial direction with the accretion rate frequently changing signs on au scale (Figure \ref{fig:accretion_rate_R} upper panel), whereas the integrated $\alpha_{int}$ has a lower variability (Figure \ref{fig:accretion_rate_R} lower panel).
For $\mathrm{r_{cav}} = 3\ \mathrm{r_\odot}$ model, the accretion rate is negative (ingoing) in the inner disk until 35 au, then it changes signs rapidly until reaching positive (outgoing) 10$^{-9}$ M$_\odot$ yr$^{-1}$ in the outer disk. Except near the inner boundary, the $\alpha_{int}$ increases from $\lesssim$ 10$^{-6}$ at 50 au to $\sim$ 10$^{-2}$ beyond 200 au. The $\mathrm{r_{cav}} = $ 18 au model has positive (outgoing) 10$^{-9}$ M$_\odot$ yr$^{-1}$ inside 25 au, but it can be affected by the setup for inner boundary. Then the accretion rate becomes negative until 100 au, and becomes positive in the outer disk. The $\alpha_{int}$ increases from $\sim$ 10$^{-4}$ to $\sim$ 10$^{-2}$ from inner to outer disk. The $\mathrm{r_{cav}} =$ 54 au model has the highest magnitude of accretion rate, reaching positive 10$^{-8}$ M$_\odot$ yr$^{-1}$ around 100 au and positive 10$^{-8}$  M$_\odot$ yr$^{-1}$ around 200 au. The low density model has almost zero accretion rate ($<$ 10$^{-11}$ M$_\odot$ yr$^{-1}$; the M$_{acc}$ for the low density model is multiplied by 100 to show its value) and $\alpha_{int}$ between $10^{-5}$-$10^{-4}$ throughout the disk. We want to emphasize that even though the fiducial \texttt{18au-rad} model has lower stress values than the equivalent vertically isothermal model \texttt{18au-rad-lowdens} at the midplane, the integrated $\alpha_{int}$ can be still larger due to the layered accretion at the disk surface.

\begin{figure*}
	\includegraphics[width=\linewidth]{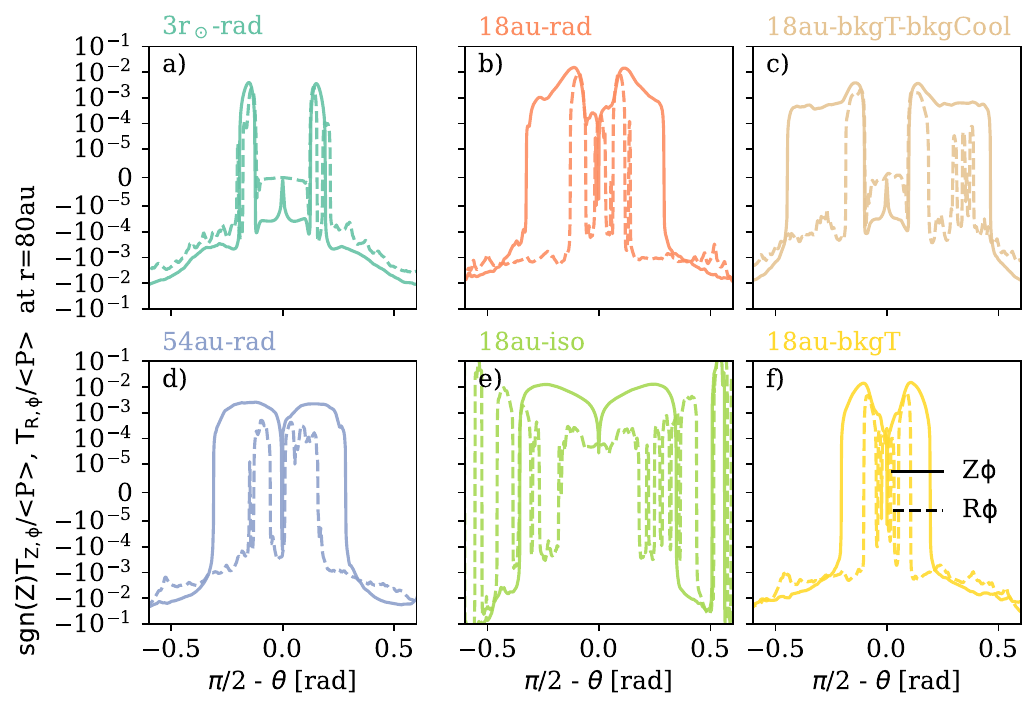}
    \caption{The turbulence stresses for Z-$\phi$ and (solid lines) R-$\phi$ (dashed lines) components normalized by the averaged gas pressure at 80 au along the vertical direction. The values are calculated between t = 1000-1200 P$_{\mathrm{in}}$, and averaged from r = 60-100 au. From left to right, they are \texttt{3r$_\odot$-rad}, \texttt{18au-rad}, \texttt{18au-bkgT-bkgCool}, \texttt{54au-rad},  \texttt{18au-iso}, and \texttt{18au-bkgT} models, respectively.}
    \label{fig:stresses_vertical}
\end{figure*}

We quantify the stress levels in Figure \ref{fig:stresses_vertical}. These stresses exhibit fluctuations with respect to time, radius, and $\theta$, thus we compute the average of these quantities between 60-100 au to represent the stress at 80 au. The time average spans 200 $\mathrm{P_{in}}$, consistent with previous figures. We present the stresses for three rad-hydro simulations ($\mathrm{r_{cav}} = 3\ \mathrm{r_\odot}$, 18 au, and 54 au models in panels a, b and d), alongside three pure-hydro simulations with various assumptions (panels c, e, and f, which we will discuss more in the next section, Section \ref{sec:approximation}). Solid lines represent $\alpha_Z$, while dashed lines represent $\alpha_R$. 

For the full disk model (\texttt{3$\mathrm{r_\odot}$-rad}), both $\alpha_R$ and $\alpha_Z$ peak around 0.2 radians, coinciding with the location of the strongest shear and reaching values on the order of $10^{-2}$. In the midplane, $\alpha_R$ remains less than $10^{-6}$, whereas $\alpha_Z$ is around $10^{-4}$. The \texttt{18au-rad} model exhibits similar behavior, with $\alpha_Z$ and $\alpha_R$ both reaching values of approximately $10^{-2}$ at the transition region between the midplane and the atmosphere. For both components, $\alpha$ remains around $10^{-4}$ to $10^{-3}$ in the midplane. In the case of the \texttt{54au-rad} model, vertical turbulence $\alpha_Z$ consistently remains at a higher level, around $10^{-3}$ to 0.4 radians, whereas $\alpha_R$ can be lower by an order of magnitude, approximately $10^{-3}$. For classical locally-isothermal and vertically isothermal VSI (\texttt{18au-iso}), the anisotropy between these two components are more pronounced, with $\alpha_Z$ ($\sim 10^{-2}$) exceeding $\alpha_R$ ($\sim 10^{-4}$) by two orders of magnitude, consistent with previous studies \citep[e.g.,][]{stoll17a}.

\begin{figure*}
	\includegraphics[width=\linewidth]{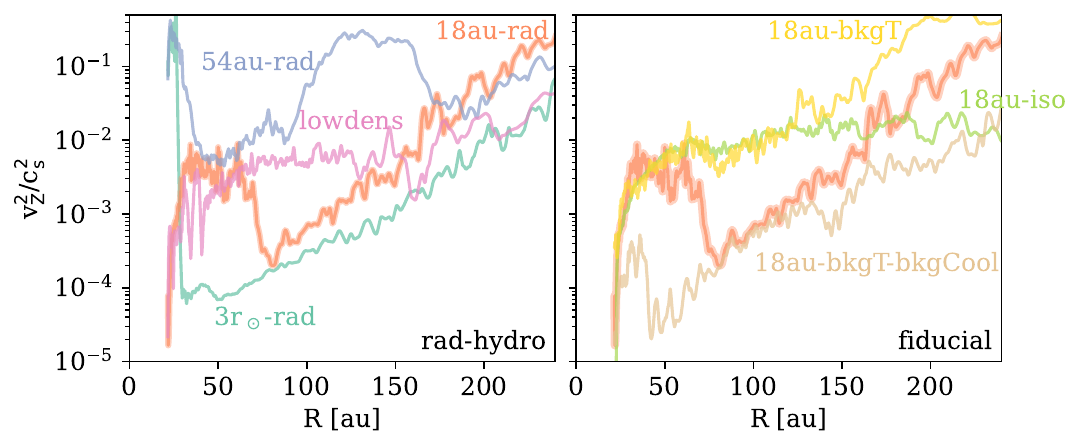}
    \caption{Time-averaged $v_Z^2$/$c_s^2$ that represents the turbulence level in the midplane for rad hydro models (\texttt{3r$_\odot$-rad}, \texttt{18au-rad},\texttt{54au-rad}, and  \texttt{18au-lowdens-rad}) on the left panel and fiducial models (\texttt{18au-rad}, \texttt{18au-iso}, \texttt{18au-bkgT}, and  \texttt{18au-bkgT-bkgCool}) on the right panel.}
    \label{fig:midplane_turbulence}
\end{figure*}

In the radial direction, not only the integrated turbulence $\alpha_{int}$ increases with radius as shown in Figure \ref{fig:accretion_rate_R}, but also the midplane turbulence. This can be seen in 2D velocity maps such as Figures \ref{fig:moneyplot} and \ref{fig:2DrhoTvZ}, and turbulence map in Figure \ref{fig:2Dshear_vrms}. We quantify this by plotting the time-averaged $v_\mathrm{Z}^2$ in Figure \ref{fig:midplane_turbulence}. On the left panel we show four rad-hydro models. For \texttt{3r$_\odot$-rad} and \texttt{18au-rad}, the vertical velocities are low in the inner disk $\sim$ 1\% of the local sound speed at $\sim$ 50 au and increases to more than 10\% of the sound speed in the outer disk at $\sim$ 200 au. For the \texttt{54au-rad} model, the turbulence level is high across the disk ($\gtrsim$ 10\% of the sound speed), where the perturbation is aligned with the zonal flow. The trend can be explained by the fact that as the effective scale height becomes larger in the outer disk, the density contrast between the midplane and atmosphere becomes smaller. At $\sim$ 200 au, the turbulent flow at the atmosphere already has enough energy to disturb the quiet midplane. For the \texttt{18au-lowdens-rad} model, the turbulence is almost constant across the disk, on the order of 10\% of the sound speed, since there is no thermal stratification.

In a recent paper by \citet{melonfuksman23a}, their $f_\mathrm{dg}$ = 10$^{-4}$ model ($f_\mathrm{dg}$ is the dust to gas mass ratio) also has a similar temperature and cooling time stratification to our fiducial model, which leads to a quiet midplane and turbulent atmosphere. We find consistent results except a higher level of turblence in the midplane. Their midplane turbulence is below 10$^{-7}$ whereas ours is between $10^{-5}$-10$^{-4}$. This difference might be explained by multiple factors. First, their inner disk has a lower $h/r$, and a lower $h/r$ leads to a lower turbulence \citep{manger20, manger21}. In Figure \ref{fig:midplane_turbulence}, we also show that the turbulence increases with $R$ since the energy contrast between the midplane and atmosphere becomes lower, which points to a lower turbulence value for the inner disk. They also have 200 cells per scale height resolution which can better resolve lower turbulence levels.

\subsection{Good Approximation: Background Temperature with Local Orbital Cooling}
\label{sec:approximation}
\begin{figure*}
	\includegraphics[width=\linewidth]{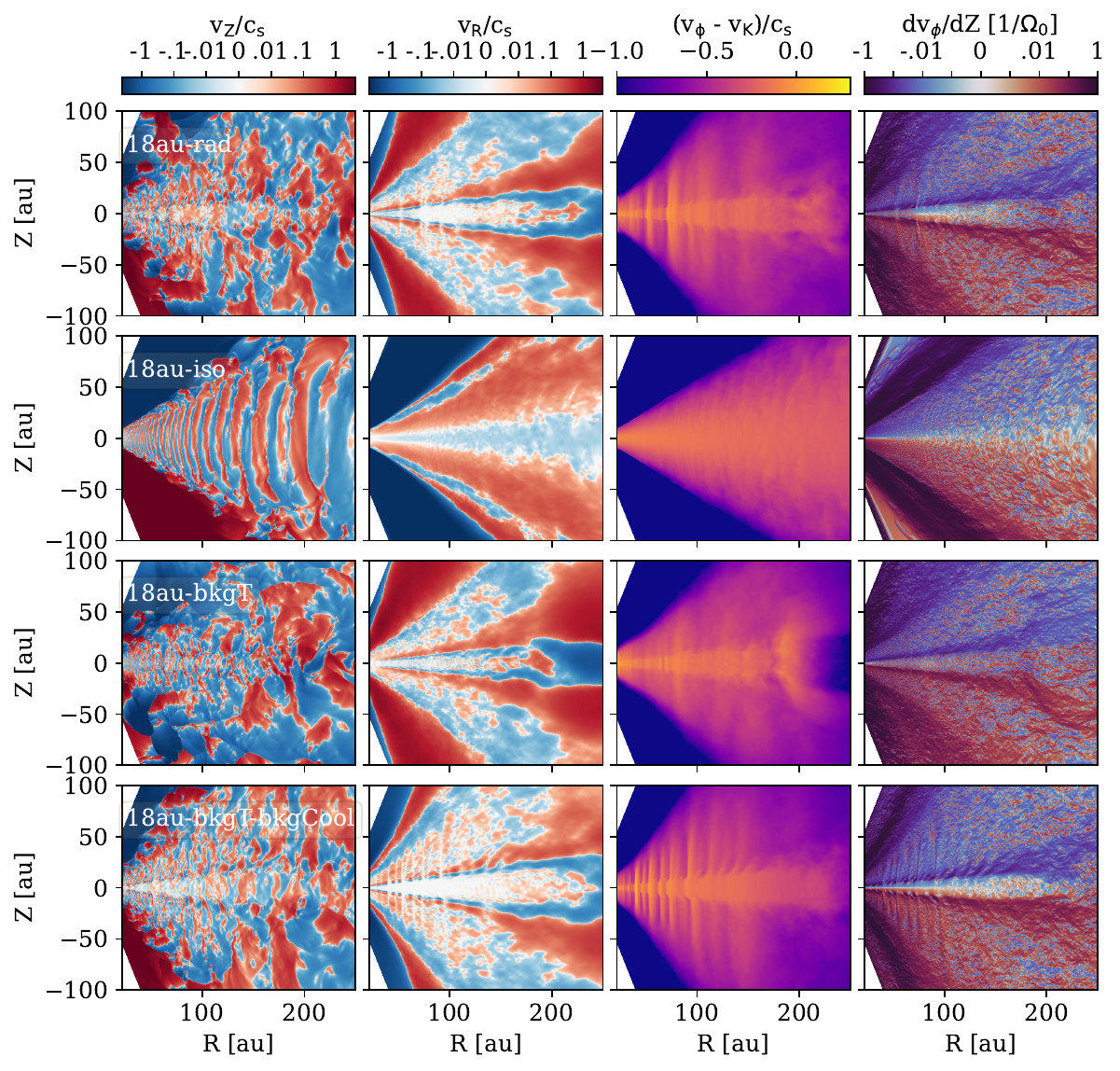}
    \caption{Comparison of the velocity fields between rad-hydro (first panel), isothermal (second panel), background temperature (third panel), and background temperature with cooling (fourth panel) models. From left to right: vertical velocity (v$\mathrm{_Z}$) snapshot, time-averaged  radial velocity (v$\mathrm{_R}$), azimuthal velocity subtracted by Keplerian velocity (v$_\mathrm{\phi}$ - v$_\mathrm{K}$), and  vertical shear rate (dv$_\mathrm{\phi}$/dZ).}
    \label{fig:compareradbkg2D}
\end{figure*}

As rad-hydro simulations differ significantly from classical VSI shown in vertically and locally isothermal simulations, we attempt to use pure-hydro simulations with varying levels of assumptions for comparison with rad-hydro simulations. We use Figure \ref{fig:compareradbkg2D} to illustrate that pure hydro simulations with background temperature and local orbital cooling provide a good approximation for rad-hydro simulations. 

Figure \ref{fig:compareradbkg2D} displays (from left to right) vertical velocity snapshots, time-averaged radial velocity, azimuthal velocity subtracted from Keplerian velocity, and vertical shear rate. From top to bottom panels, we have the rad-hydro fiducial model (\texttt{18au-rad}), the vertically and locally isothermal model (\texttt{18au-iso}), the locally isothermal and background temperature model (\texttt{18au-bkgT}), and the orbital cooling and background temperature model (\texttt{18au-bkgT-bkgCool}). The rad-hydro model has been introduced in Figures \ref{fig:2DrhoTvZ} and \ref{fig:2DvRvphidphidZ}; we retain them since these panels can be directly compared with the isothermal model in the second row, representing a classical VSI picture. The isothermal model (\texttt{18au-iso}) also resembles the low-density rad-hydro model. We note that the radial temperature gradient and $h/r$ in \texttt{18au-iso} model are measured in the midplane of rad-hydro simulations for a close comparison. In the pure-hydro simulation that incorporates the background temperature of the rad-hydro simulation (third row, \texttt{18au-bkgT}), we observe the disruption of the delicate n=1 corrugation mode. Layered accretion and strong shear in the temperature transition region become evident, but the midplane vertical velocity remains higher than in rad-hydro simulations, as this region is still VSI unstable due to the almost zero cooling time ($\beta$ = 10$^{-6}$). The zonal flow in the azimuthal velocity also differs from the rad-hydro simulation. 

By incorporating the estimated cooling time from the rad-hydro simulation in the bottom panels (\texttt{18au-bkgT-bkgCool}), all four fields closely resemble the rad-hydro simulations. Therefore, we demonstrate that computationally inexpensive pure-hydro simulations can capture crucial features of rad-hydro simulations. In future, one can prescribe a temperature structure or obtain a self-consistent temperature structure by iterating the temperature calculated from Monte Carlo Radiative Transfer code such as RADMC-3D \citep{dullemond12} and enforcing vertically hydrostatic equilibrium \citep{bae19,ueda19,zhang21}. Then the cooling structure can be estimated using Equation \ref{eq:beta}.

It is important to note that they are not identical; an apparent difference is that the zonal flow in the pure-hydro model has narrower length scales than in the rad-hydro simulations. This discrepancy could be attributed to the static temperature and estimated cooling profiles in pure-hydro simulations, while rad-hydro simulations undergo secular evolution of temperature and cooling times. Additionally, the cooling is no longer local, since the radiative cooling can be affected by other parts of the disk. Their Reynolds stresses are similar but not identical. 

In Figure \ref{fig:stresses_vertical}, the \texttt{18au-bkgT-bkgCool} and \texttt{18au-bkgT} models exhibit shapes more similar to the rad-hydro model (\texttt{18au-rad}) but still display some differences. Their $\alpha_Z$ and $\alpha_R$ exhibit similar magnitudes, indicating that the turbulence becomes much more isotropic than in the isothermal model (\texttt{18au-iso}). However, \texttt{18au-bkgT-bkgCool} shows lower turbulence levels, and the sign of Z-$\phi$ stress differs from \texttt{18au-rad} beyond 0.3 radians and within 0.1 radians. Additionally, $\alpha_R$ is smaller in the midplane. The $\alpha_Z$ of the \texttt{18au-bkgT} model has a different sign than \texttt{18au-rad} between 0.2-0.3 radians. Overall, values from rad-hydro models and those pure-hydro models that adopted the rad-hydro thermal structures have smaller than $10^{-3}$  $\alpha_Z$ in the midplane, except in the case of a large cavity (54 au). They also exhibit more isotropic turbulence than the isothermal one. The low-density case is not presented here, but its stress profile is similar to that of an isothermal simulation with a larger $h/r$ than the \texttt{18au-iso} model due to its almost isothermal profiles and low equivalent cooling time (Figures \ref{fig:2DrhoTvZ} to Figure \ref{fig:2DvRvphidphidZ}).

In the radial direction, the turbulence levels between the pure hydro and rad-hydro simulations are similar but not identical. On the right panel of Figure \ref{fig:midplane_turbulence}, we show pure hydro models along with the fiducial rad-hydro model. \texttt{18au-bkgT} and \texttt{18au-bkgT-bkgCool} models follow the fiducial model's trend, but the former overestimates the turbulence whereas the latter underestimate the turbulence. The local orbital cooling prescription strongly under-predict the turbulence level within 70 au. The vertically isothermal model has almost constant turbulence level around the disk similar to that of the low density rad-hydro model.

\section{Discussion}
\label{sec:discussion}
\subsection{Gas Substructures}
\label{sec:substructure}
\begin{figure*}
	\includegraphics[width=\linewidth]{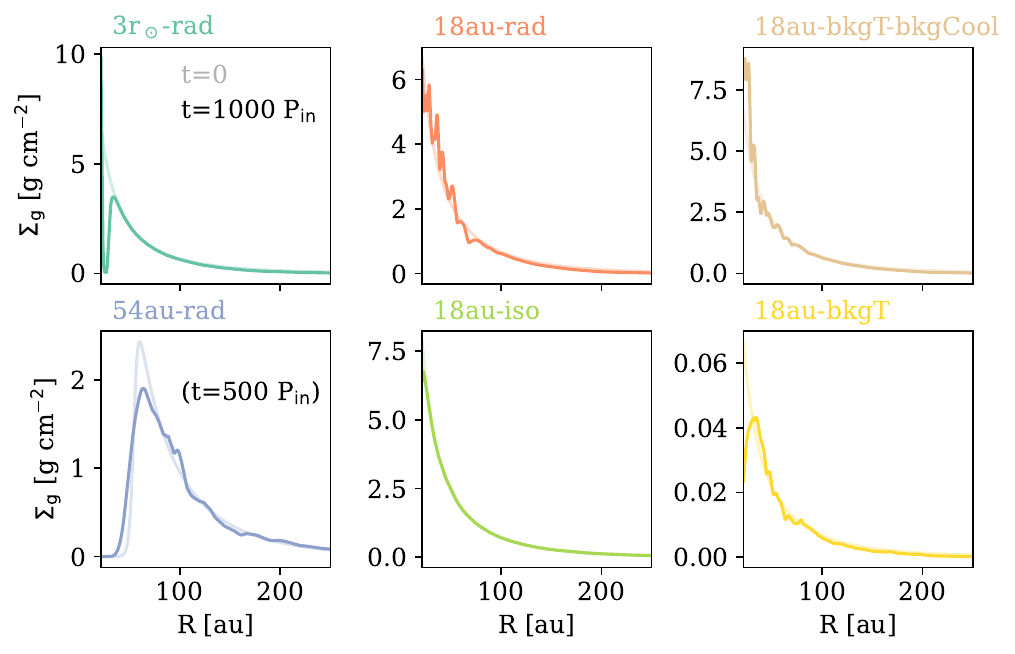}
    \caption{The gas surface density profiles in the radial direction at t = 1000 P$_\mathrm{in}$ or 500 P$_\mathrm{in}$ for \texttt{54au-rad} (more opaque lines) and at the initial condition (more transparent lines). The layout is the same as Figure \ref{fig:stresses_vertical}.}
    \label{fig:surface_density}
\end{figure*}
Related to the zonal flows shown in Figure \ref{fig:compareradbkg2D}, gas substructures can also develop depending on the inner cavity size as shown in Figure \ref{fig:surface_density}. The full disk model (\texttt{3r$_\odot$-rad}) preserves the initial condition except at the inner disk due to the boundary effect. The \texttt{18au-rad} model has perturbation on several au scale, whereas the \texttt{54au-rad} model has perturbation on tens of au scale. \texttt{18au-bkgT-bkgCool} and \texttt{18au-bkgT} models have similar perturbations as the \texttt{18au-rad}, whereas the isothermal model \texttt{18au-iso} keeps the initial condition. 

\begin{figure*}
	\includegraphics[width=\linewidth]{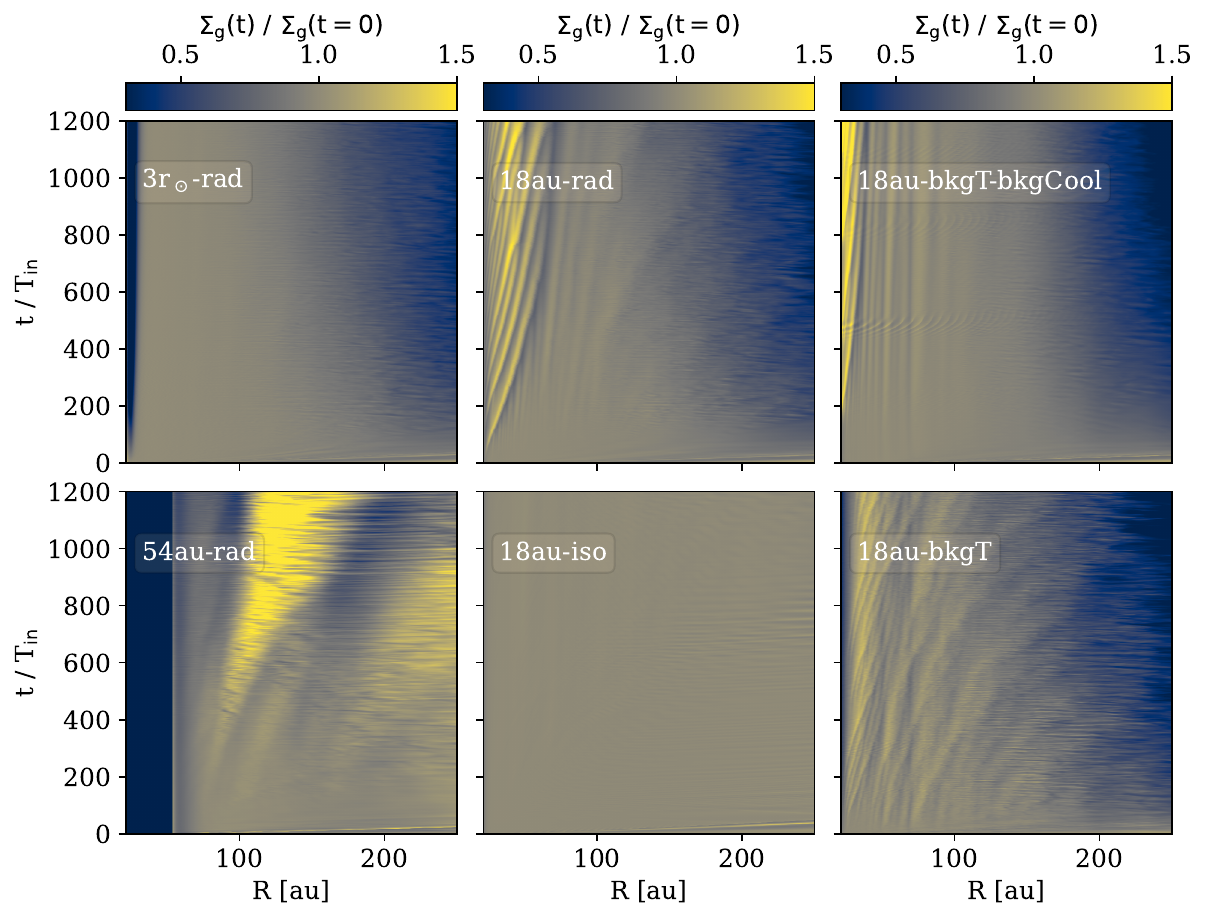}
    \caption{The gas surface density normalized by the initial condition as a time evolution from t = 0-1200 P$_\mathrm{in}$. Brown colors mean that the surface density is almost unchanged. Yellow colors indicate the increase of density, whereas blue colors indicate the decrease of density. The layout is the same as Figure \ref{fig:stresses_vertical}.}
    \label{fig:surface_density_evolution}
\end{figure*}

Figure \ref{fig:surface_density_evolution} shows the time evolution of the surface density. The full disk model (\texttt{3r$_\odot$-rad}) and the isothermal model (\texttt{18au-iso}) do not show evident substructures. For the rest of the models, substructures can form and propagate to the outer disk. for $\mathrm{r_{cav}}$ = 18 au models, some of these rings form at the inner boundary, but other rings can form in the middle of the disk and move to the outer disk at a lower speed. The time evolutions between \texttt{18au-rad}, \texttt{18au-bkgT}, and \texttt{18au-bkgT-bkgCool} are not identical. The rings in \texttt{18au-bkgT-bkgCool} model move at a lower speed than those in \texttt{18au-rad}.  For the \texttt{54au-rad} model, the perturbation becomes much stronger after 700 orbits due to the vortex formation around 100 au in R-$\theta$ plane. However, in 3D simulations, vortices tend to develop in R-$\phi$ plane. Future 3D studies will unveil a more realistic structure of this model. These gas substructures can also possibly lead to dust substructures, but this needs to be tested in future studies that includes dust particles. Overall, our limited sample of rad-hydro models shows a tentative trend that transition disks with larger cavity sizes are more prone to develop zonal flows and substructures.

\subsection{Observational and Modeling Prospect}
\label{sec:observation}
\begin{figure*}
 \includegraphics[width=\linewidth]{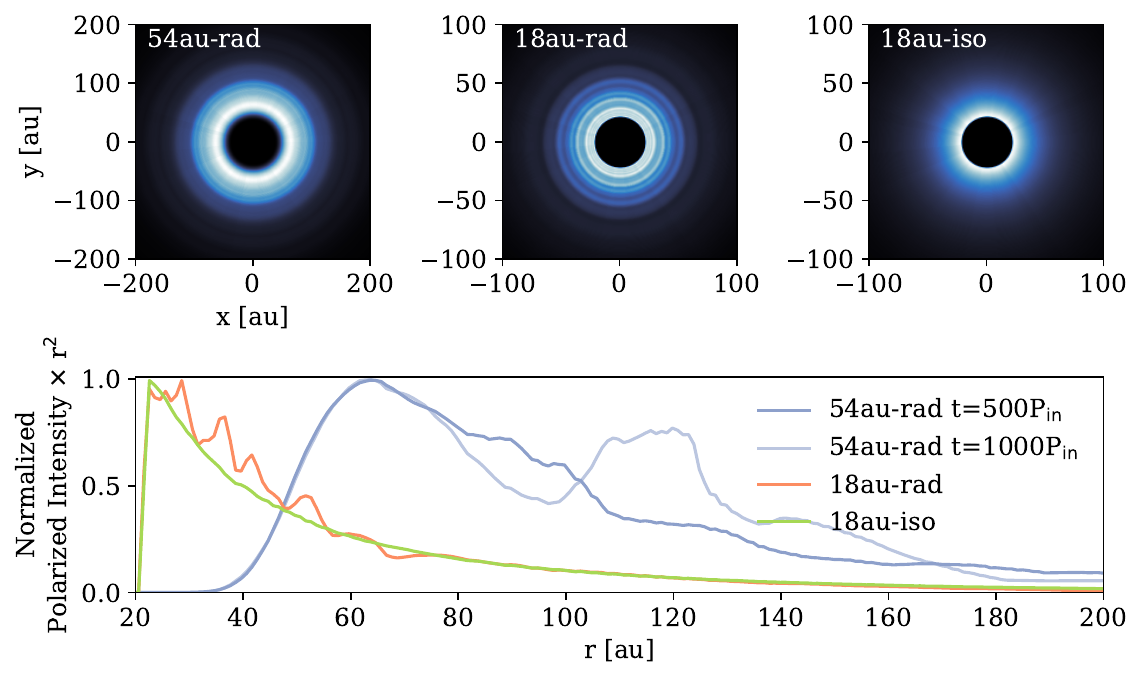}
    \caption{Near infrared scattered light images (polarized intensity) at H-band (1.63 $\mu$m) for \texttt{54au-rad}, \texttt{18au-rad}, and \texttt{18au-iso} models.}
    \label{fig:scatteredlight}
\end{figure*}
While we focus on 2D modeling, the thermal structure is calculated self-consistently so we can make some predictions for axis-symmetric disks. In Figure \ref{fig:scatteredlight} we show the near-infrared scattered light polarized intensity at $\lambda$=1.63 $\mu$m (H-band) using RADMC-3D for face-on \texttt{54au-rad}, \texttt{18au-rad}, and \texttt{18au-iso} models in linear scale from 0 to maximum value. We used the same DSHARP opacity (Figure \ref{fig:opac}), assuming that the dust to gas ratio is 0.01 and small grains (0.1-1 $\mu$m) account for 0.02184 of the total dust mass. The temperature is directly taken from rad-hydro simulations. For the isothermal model (\texttt{18au-iso}), we run thermal Monte Carlo (\texttt{radmc3d} \texttt{mctherm}) to calculate temperature in the $r-\theta$ plane. The RADMC-3D calculated the full Stokes image using the scattering matrix. The polarized intensity is $(Q^2+U^2)^{1/2}$. For the large cavity transition disk model \texttt{54au-rad} (taken at t = 500 P$_\mathrm{in}$), we can clearly observe the inner rim and rings at around 100 au. If we take the snapshot at a later time, the ring structure becomes more evident, consistent with the surface density perturbation in Figure \ref{fig:surface_density_evolution}. However, we need to test whether this perturbation is still large in 3D simulations in future. For \texttt{18au-rad} model, several rings can also be seen close to the inner rim, resembling Figure \ref{fig:surface_density}, but the length scale is smaller than the disk with a larger cavity, making their substructures more difficult to be observed. In contrast, we can only see the inner rim of the vertically and locally isothermal model (\texttt{18au-iso}), meaning that the outer disk is in the shadow \citep[i.e., this is a self-shadowed disk as defined in][]{garufi18,garufi22}. While we only have limited numbers of rad-hydro models with varying cavity sizes, we find the tendency that disks with larger cavity sizes can produce wider rings and can be easier to be observed in near-infrared scattered light images. This is consistent with the finding in current scattered light disk demographics which shows that ring structures are predominantly found in disks with weak Near-IR excess \citep{benisty22}, where weak Near-IR excess is often interpreted as no inner disk.  

As shown in Figure \ref{fig:midplane_turbulence}, the turbulence level in the vertical direction appears to be stronger in the region where the disk is directly exposed to stellar irradiation. For example, \texttt{54au-rad} and \texttt{18au-rad-lowdens} models that have less attenuation of the stellar irradiation have higher turbulence values than the \texttt{3$_\odot$-rad} and \texttt{18au-rad} models.
This suggests that dust turbulent diffusion could be also strong at the cavity edge. In HD 163296, \citet{rosotti20, doi21, doi23} have found that the $\alpha$/St value is higher for the inner ring at 68 au, which is more exposed to stellar irradiation than the outer ring at 101 au. This finding aligns with our results, assuming that the Stokes number (St) does not vary significantly between these two rings. To validate the impact of direct stellar irradiation on dust diffusion, it will be crucial to measure dust settling at the cavity edges through ALMA observations and incorporate dust particles into 3D hydro simulations. Furthermore, in Figure \ref{fig:midplane_turbulence}, for full disks or transition disks with small cavities, midplane turbulence values increase with radius, a trend that can be readily examined through ALMA observations.

\begin{figure*}
 \includegraphics[width=\linewidth]{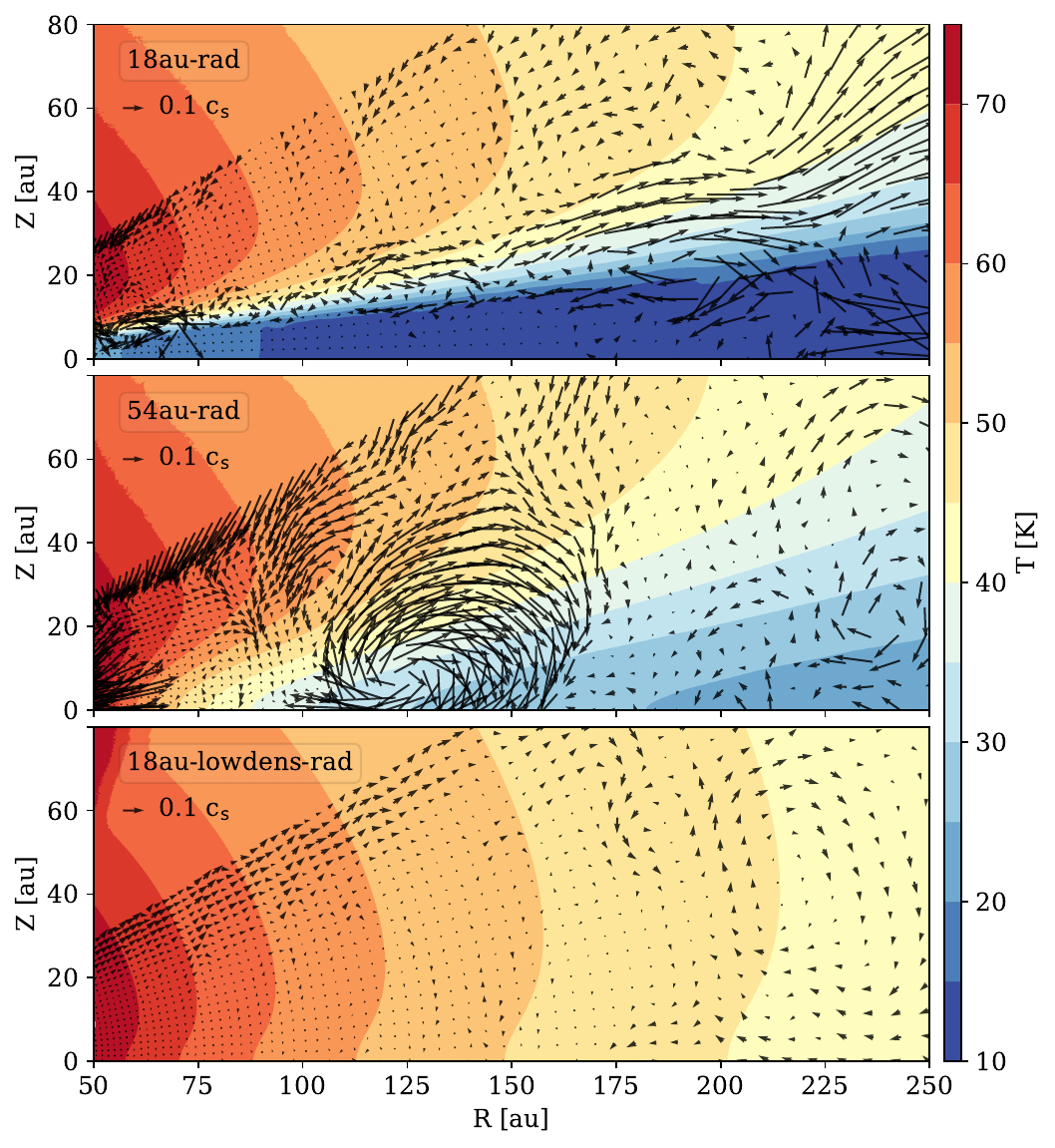}
    \caption{Time-averaged temperature and meridional velocity vectors for \texttt{18au-rad}, \texttt{54au-rad}, and \texttt{18au-lowdens-rad} models. 10\% of the local sound speed is shown on the upper left of each panel.}
    \label{fig:quiver_plot}
\end{figure*}

Different molecular lines from the ALMA MAPS Large Program \citep{oberg21b} and other high-resolution, high-sensitivity datasets are used to map the temperature structure in protoplanetary disks \citep{zhangk21, law21, law22, law23, law24}. Conversely, these data can also reveal the velocity vectors at line emission surfaces, as indicated in numerous studies \citep[see a review by ][]{pinte22}. Specifically, \citet{teague19, yu21, galloway-sprietsma23} identify disk winds and meridional flows attributed to embedded planets. Currently, we anticipate a substantial increase in sample size from the ALMA Large Program exoALMA. Aligning with the theme of the current paper--where temperature structure influences kinematics--we use Figure \ref{fig:quiver_plot} to illustrate our ability to establish correspondence between the temperature structure and kinematic features. The figure shows temperature contours overlaid by meridional velocity vectors for the \texttt{18au-rad}, \texttt{54au-rad}, and \texttt{18au-lowdens-rad} models. In our fiducial \texttt{18au-rad} model, the temperature is 10-20 K in the midplane, and the velocity remains low within 200 au. At the transition region from the midplane to the atmosphere, the gas flows inward, reaching 10\% of the sound speed. At higher altitudes, the gas flows outward with increasing velocity towards the outer disk. In contrast, for the low-density model, since the temperature has no vertical dependence, the gas velocity remains low throughout the disk. The change of the flow structure in response to the temperature structure can be identified in observations. However, the emission surface does not necessarily parallel the streamline. For example, if the emission surface height increases with radius first and then decreases due to a lower optical depth (e.g., Figure 4 in \citealt{galloway-sprietsma23}), changing directions of the velocity measured along the emission surface can be simply attributed to the multiple crossings of the streamlines, without the need of any radial substructure, such as disk wind or meridional flow. A recent study by \citet{martire24} has demonstrated that the difference in inferred rotational velocities from $^{12}$CO and $^{13}$CO can be attributed to vertical thermal stratification (see middle panels of Figure \ref{fig:2DvRvphidphidZ}), with the midplane (traced by $^{13}$CO) exhibiting a higher rotational velocity than the atmosphere (traced by the more optically thick $^{12}$CO). By accounting for thermal stratification, they can retrieve properties such as disk mass and stellar mass more accurately. In our current paper, we want to highlight that considering thermal structure can lead to further consequences, including spatially varying radial velocity, vertical velocity, and zonal flow.

The middle panel of Figure \ref{fig:quiver_plot} shows the temperature and flow structure of the transition disk \texttt{54au-rad}. In this case, the temperature varies both vertically and radially. At the cavity edge (Figure \ref{eq:temperature}, 50-100 au), the radial temperature gradient is steeper than that in a smooth disk. A giant clockwise rotating vortex is also present. While the existence and strength of the vortex needs to be tested in 3D simulations, it could result from the radial and vertical variation of the temperature instead of an artifact from the rad-hydro simulation, since a similar feature was also found in a pure-hydro simulation, \texttt{54au-bkgT-bkgCool}. Future efforts will focus on validating the existence and possible forming mechanism of this vortex through dedicated 3D simulations. We note that the temperature variation across the transition disk cavity has been identified from observations \citep{leemker22} and can shape the variety of transition disk gas substructures \citep{wolfer23}, meaning that the correspondence between the temperature structure and the kinematic features in transition disk can also be probed with current observations. 

It has been well-known that temperature structure strongly shapes disk chemistry, but we also aim to use Figure \ref{fig:quiver_plot} to demonstrate that the thermal structure can influence disk chemistry by shaping disk kinematics, thereby affecting material transport. While static disk models cover more complete chemical networks, studies considering dynamical effects \citep[e.g., ][]{aikawa99, semenov11, furura13, furura14, price20, bergner21, vanclepper22} by incorporating radial and/or vertical gas/dust mixing often reveal different chemical distributions compared to static models. These dynamic models may provide a better explanation for observations \citep[see reviews by][and references therein]{krijt22, oberg23}. We anticipate that, by accounting for advection and turbulent diffusion due to a specific thermal structure, the chemical distribution can be more accurately predicted. For instance, the layered accretion in our fiducial model \texttt{18au-rad}, where an outgoing flow is atop an ingoing flow, can transport material inward in the colder layer and outward in the hotter layer, which may also have implications on solid transport in our solar system \citep{ciesla09}.
The turbulence level measured at both the midplane and the atmosphere in our fiducial model has some similarities to layered accretion models featuring an MRI-active atmosphere and a dead zone in the midplane \citep{gammie96, simon11, simon13, bai15, xu17, simon18}. On the other hand, the high turbulence values in the upper layer are only measured in a few of protoplanetary disks \citep{flaherty20, paneque-carreno23} (see a recent review by \citealt{rosotti23}). Our objective is to conduct 3D simulations to predict unique channel map features, following previous studies by \citet{hall20, barraza-alfaro21}. As demonstrated in Section \ref{sec:approximation} (Figure \ref{fig:compareradbkg2D}), background temperature and local orbital cooling profiles prove to be effective approximations for rad-hydro simulations. This sets the groundwork for using 3D simulations in our future work to make ALMA and near-infrared scattered light observational predictions.

\section{Conclusions}
\label{sec:conclusions}
Vertical shear instability (VSI) is a promising candidate to generate turbulence in the outer region of protoplanetary disks. It can be crucial for gas and dust transport in protoplanetary disks. We study VSI using the Athena++ radiation module with stellar irradiation, which self-consistently captured the thermal structure and hydrodynamics. We study disks with different inner cavity sizes, accompanied by pure hydro simulations with various assumptions. We find that temperature structure strongly influences disk kinematics. Our main findings are as follows:
\begin{enumerate}
  \item The radial optical depth of the star determines the disk's thermal structure. For realistic disk setups (M$_d$ = 10$^{-2}$ M$_\odot$, cavity size $<$ 54 au), the disk can be separated into the cool midplane and super-heated atmosphere, delineated by the $\tau_*$ = 1 surface (Figure \ref{fig:2DrhoTvZ}). If the disk is optically thin to the stellar irradiation (low mass disk with M$_d$ = 10$^{-4}$ M$_\odot$), the temperature is almost vertically isothermal.
  \item  The thermal structure determines disk's kinematics. The temperature and cooling time ($\beta$) stratification suppresses the classical n = 1 corrugation mode that leads to meridional circulations found in isothermal simulations (Figure \ref{fig:moneyplot}). Instead, the turbulence becomes more isotropic on a more local scale, in contrast to very large vertical-azimuthal Reynolds stress $\alpha_Z$ ($\sim 10^{-2}$) and weak radial-azimuthal Reynolds stress $\alpha_R$ ($\sim 10^{-4}$) found in isothermal simulations (Figure \ref{fig:stresses_vertical}). The low mass disk has vertically isothermal profiles, so it closely resembles all features in isothermal simulations.
  \item The strongest vertical shear occurs at the transition region between the cool midplane and superheated atmosphere where many vortices form. At this transition region, layered accretion happens with an outgoing flow on top of an ingoing flow (Figure \ref{fig:2DvRvphidphidZ}). This layered accretion can be perfectly explained by the vertical variation of the stress structure (Figure \ref{fig:2Dshear_vrms}) using Equation \ref{eq:amephi2}.
  \item Pure hydro simulations with measured temperature structures and estimated orbital cooling profiles can be good approximations for rad-hydro simulations (Figure \ref{fig:compareradbkg2D}).
  \item Zonal flows and gas substructures can develop, and a disk with a larger cavity size has perturbations with a longer length scale and stronger magnitude (Figures \ref{fig:2DvRvphidphidZ} and \ref{fig:surface_density}). At the cavity edge, the gas has stronger turbulence, which could slow dust settling. Using MCRT simulations, we confirm that transition disks tend to have rings, and the disks with larger cavities tend to have more prominent rings, which are easier to be observed in near-infrared scattered light images (Figure \ref{fig:scatteredlight}), consistent with the fact that rings in scattered light images are predominantly found in disks with weak Near-IR excess.
\end{enumerate}

We also show the correspondence between the temperature structure and kinematic features (Figure \ref{fig:quiver_plot}). Future work including synthetic observations can be used to predict this correspondence in ALMA observations. The temperature structure can also influence chemistry through shaping the disk kinematics.  One can also predict the observational signatures in ALMA chemistry observations by modeling disk chemistry with disk dynamics under such a thermal structure.

\section*{Acknowledgments}
We thank the anonymous referee's careful review and constructive comments. S.Z. thanks Chao-Chin Yang, Kaitlin Kratter, Leonardo Krapp, Andrew Youdin, and Pinghui Huang for helpful discussions.
All simulations are carried out using computers from the NASA High-End Computing (HEC) program through the NASA Advanced Supercomputing (NAS) Division at Ames Research Center. S.Z. and Z.Z. acknowledge support through the NASA FINESST grant 80NSSC20K1376. S.Z. acknowledges support from Russell L. and Brenda Frank Scholarship. Z. Z. acknowledges support from the National Science Foundation under CAREER grant AST-1753168 and support from NASA award 80NSSC22K1413. The Center for Computational Astrophysics at the Flatiron Institute is supported by the Simons Foundation.

\vspace{5mm}

\software{Athena++ \citep{stone20}, RADMC-3D \citep{dullemond12}, 
          Matplotlib \citep{hunter07}, SciPy \citep{scipy20}, lic (\url{https://gitlab.com/szs/lic})
          }

\appendix

\section{Implementation of Stellar Irradiation and Unit Conversion}
\label{sec:raytracing}

We employ the Athena++ frequency-integrated implicit radiation module \citep{jiang21} to solve the following set of equations: three hydrodynamic equations and one radiation transfer equation. Additionally, in the energy equation, we incorporate stellar irradiation as a source term (- $\bfnabla\cdot \bF_*$), similar to \citet{flock17}.

The set of equations is
\begin{align}
\frac{\partial\rho}{\partial t}+\bfnabla\cdot(\rho \bvp)&=0, \nonumber \\
\frac{\partial( \rho\bvp)}{\partial t}+\bfnabla\cdot({\rho \bvp\bvp +{{P\mathbf{I}}}}) &=-\bm{S_r}(\bP) + \rho \mathbf{a_{\mathrm{grav}}},\  \nonumber \\
\frac{\partial{E}}{\partial t}+\bfnabla\cdot\left[(E+P)\bvp\right]&=-S_r(E) - \bfnabla\cdot \bF_* + \rho \mathbf{a_{\mathrm{grav}}} \cdot \bvp,  \nonumber \\
\frac{\partial I}{\partial t}+c\bn\cdot\bfnabla I&=cS_I,
\label{eq:hd}
\end{align}

Here, $\rho$ represents the gas density, $\bvp$ is the flow velocity, $P$ denotes the gas pressure, $\mathbf{I}$ is the unit tensor, $E$ is the total energy, $I$ represents the lab-frame specific intensity of photons emitted by the disk locally, $c$ is the speed of light, and $\bn$ is the angle in the lab frame. The terms $\bm{S_r}(\bP)$ and $S_r(E)$ represent the disk's radiation source terms in the momentum and energy equations. They are moments of the source term $S_I$ in the radiation transfer equation, given by:

\begin{align}
S_I\equiv \Gamma^{-3}\left[\right. \rho(\kappa_s + \kappa_a)\left(J_0-I_0\right)\nonumber\\
+\rho\kappa_{P}\left(\frac{a_rT^4}{4\pi}-J_0\right)\left.\right],\nonumber\\
S_r(E)\equiv 4\pi c\int S_I d\Omega,\nonumber\\
\bm{S_r}(\bP)\equiv 4\pi \int \bn S_I d\Omega,
\end{align}

where $\kappa_a$, $\kappa_s$, and $\kappa_{P}$ are the Rosseland mean opacity, scattering opacity, and Planck mean opacity, respectively. These opacities are all normalized to the gas. The intensity $I$ in the lab frame is related to the intensity in the co-moving frame $I_0$ through a Lorentz transformation:

\begin{eqnarray}
I_0(\bn^{\prime})=\gamma^4\left(1-\bn\cdot\bvp/c\right)^4 I(\bn)\equiv \Gamma^4(\bn,\bvp)I(\bn),
\label{eq:lorentz_transform}
\end{eqnarray}

where $\gamma\equiv 1/\sqrt{1-v^2/c^2}$ is the Lorentz factor, $\Gamma(\bn,\bvp)\equiv \gamma\left(1-\bn\cdot\bvp/c\right)$, and $\bn^{\prime}$ is the angle in the co-moving frame given by:

\begin{eqnarray}
\bn^{\prime}=\frac{1}{\gamma\left(1-\bn\cdot\bvp
/c\right)}\left[\bn-\gamma\frac{\bvp}{c}\left(1-\frac{\gamma}{\gamma+1}\frac{\bn\cdot\bvp}{c}\right)\right].
\label{eqn:cm_ang}
\end{eqnarray}

The angular-averaged mean intensity in the co-moving frame $J_0$ is defined as:

\begin{eqnarray}
J_0\equiv \frac{1}{4\pi}\int I_0d \Omega_0,
\end{eqnarray}

where $\Omega_0$ represents the angular element in the co-moving frame.

The total gas energy density $E$ is given by:

\begin{eqnarray}
E=E_g+\frac{1}{2}\rho v^2,
\end{eqnarray}

where $E_g$ represents the gas internal energy. Assuming an ideal gas equation of state (EoS), the internal energy is related to the gas pressure $P$ through the adiabatic index $\gamma_g$ as $E_g=P/(\gamma_g-1)$ for $\gamma_g \neq 1$. The gas temperature $T$ is calculated using $T=\mu P/\left(R_{\text{ideal}}\rho\right)$, where $R_{\text{ideal}}$ is the ideal gas constant and $\mu$ is the mean molecular weight. We adopted $\gamma_g$ = 1.4 and $\mu$ = 2.3 in this paper.

We adopted temperature, density and length units to be $T_0,\rho_0,r_0$ respectively. The time unit is given by $\Omega_{K,0}^{-1}$ = $(GM_* r_0^{-3})^{-1/2}$. The velocity unit $v_0$ is then the Keplerian velocity at $r_0$. These parameters are used to calculate two key parameters in the radiation module $\Prat\equiv a_rT_0^4/\left(\rho_0 R_{\text{ideal}}T_0/\mu\right)$ and $\Crat\equiv c/a_0$ \citep{jiang12}. $a_0$ is the characteristic isothermal sound speed, $\big(R_{\mathrm{ideal}}T_0/\mu \big)^{1/2}$. The values for our fiducial model are $T_0$ = 6.14 $\times$ 10$^{3}$ K; $\rho_0$ = 4.28 $\times$ 10$^{-14}$ g cm$^{-3}$; r$_0$ = 40 au; hence $\Prat$ =  1.13 $\times$ 10$^{3}$; and $\Crat$ = 6.36 $\times$ 10$^{4}$. $\Prat$ represents the ratio between the radiation pressure and gas pressure at these unit quantities, whereas $\Crat$ represents the ratio between speed of light and characteristic sound speed $a_0$. The $\Prat$ is larger than unity since we adopt a very large value of $T_0$. We adopt this exact value of $T_0$ since $v_0 = a_0$ so we do not need to distinguish between the unit velocity and the characteristic sound speed. For typical values of density ($\sim$ $\rho_0$) and temperature (tens of Kelvins) at the midplane, the radiation pressure is much less than the gas pressure. Similarly, the typical ratio between the speed of light and local sound speed is  $\gg\Crat$, as $T_0$ is much greater than a typical disk temperature. 
We set density floor to be $10^{-12}$ and pressure floor to be $10^{-15}$ in code units. We also set a temperature floor to be 0.001 $T_0$ (6.14 K) and a temperature ceiling to be 0.1 $T_0$ (614 K) to avoid numerical hotspots.

To account for stellar irradiation, we include a heating source term in the energy equation. This source term is necessary for frequency-integrated radiation transport \citep[e.g., ][]{flock17} as the stellar irradiation is at significantly higher temperatures (thousands of Kelvins) compared to the thermal emission from the disk (tens to hundreds of Kelvins). In a future work (Baronett et al., in prep) that uses a multi-group radiation module \citep{jiang22}, this source term is not required, and it can better capture the multi-frequency nature of radiation transport both for the stellar irradiation and disk emission.

The stellar irradiation heating flux $\mathbf{F}_*(r)$ is given by:
\begin{equation}
\rm \mathbf{F}_*(r) = \left (  \frac{R_*}{r}\right )^2  \sigma_b T_*^4 e^{-\tau} \hat{\mathbf{r}}, 
\label{eq:IRRAD}
\end{equation}
where $\rm T_*$ and $\rm R_*$ represent the stellar surface temperature and radius, respectively. Here, $\rm \sigma_b$ denotes the Stefan-Boltzmann constant, which is related to the radiation constant $a_r$ as $\sigma_b = a_r c/4$. The radial optical depth for the star at each $\theta$ is given by:
\begin{align}
\rm \tau_*(r,\theta)=\int_{R_*}^r \kappa(T_*, \theta) \rho_{dust}(r,\theta)  dr \nonumber\\ = \rm \tau_{*,bc}(r,\theta) + \tau_{*,domain}(r,\theta) \nonumber\\
\rm = \int_{R_*}^{r_{in}} \kappa(T_*, \theta) \rho_{dust}(r,\theta)  dr + \int_{r_{in}}^r\kappa(T_*, \theta) \rho_{dust}(r,\theta)  dr \, ,
\label{eq:TAU}
\end{align}
where $r_{\text{in}}$ represents the inner radius of the computational domain. The first term in the second line of Equation \ref{eq:TAU} refers to the optical depth within the region interior to the computational domain. These values are not evolved with time but depend on the density and opacity setup of the global disk. Namely, they depend on the inner cavities' radii, gas scale height, and surface density.

To ensure compatibility with MPI (Message Passing Interface), where the ray-tracing needs to cross all the grids in the radial direction and a ray can enter different \texttt{MeshBlock}s located on different CPUs, we adopted the following procedure. First, we calculated the optical depth within each \texttt{MeshBlock}. Then, we declared a user-defined \texttt{Mesh} data array to store all the $\tau_*$ values at the outer boundaries of each \texttt{MeshBlock}. These values represent the local optical depths integrated from the inner boundary ($\rm r_{mb,in}$) to the outer boundary ($\rm r_{mb,out}$) of each \texttt{MeshBlock}. For the zeroth column, it stores values of $\rm \tau_{*,bc}$. In the middle of each timestep, we cumulatively sum the the user defined \texttt{Mesh} data in the radial direction. Then an \texttt{MPI\_Allreduce} operation is performed to update all the user-defined \texttt{Mesh} data array in all the CPUs so that the inner boundary optical depths of each \texttt{MeshBlock} has their correct global values. Finally, the optical depth $\tau_*$ within each \texttt{MeshBlock} can be calculated by adding up its current \texttt{MeshBlock}'s inner boundary value and its local integrated value. Specifically, within each \texttt{MeshBlock}, the optical depth is given by:
\begin{align}
\rm \tau_*(r,\theta)= \tau_*(r_{mb,in},\theta) + \int_{r_{mb,in}}^r\kappa(T_*, \theta) \rho_{dust}(r,\theta)  dr , 
\end{align}
where $\rm \tau_*(r_{mb,in},\theta)$ is the global optical depth at the \texttt{MeshBlock}'s inner boundary stored in the user-defined \texttt{Mesh} data array.

\section{Energy Budget}
\begin{figure*}
	\includegraphics[width=\linewidth]{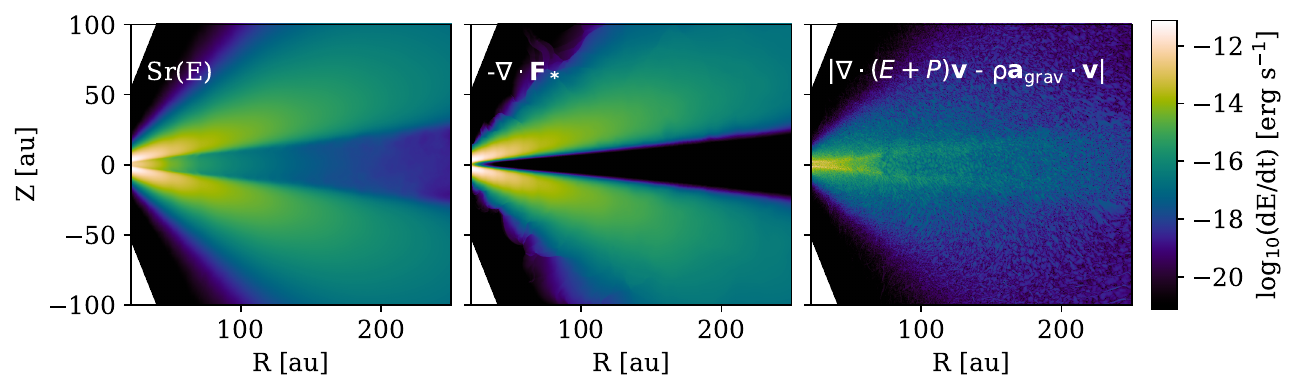}
    \caption{Time averaged (t=1000-1200 P$\mathrm{_{in}}$) source terms and energy flux divergence in the energy equation (Equation \ref{eq:hd}) for the fiducial model \texttt{18au-rad}. Left panel: disk's radiation source term ($Sr(E)$); middle panel: stellar irradiation heating source term (-$\bfnabla\cdot \bF_*$); right panel:  energy flux divergence including gravitational energy.}
    \label{fig:energy_budget}
\end{figure*}

We use the energy equation in Equation \ref{eq:hd} and refer to Figure \ref{fig:energy_budget} to illustrate the energy budget in our fiducial model, \texttt{18au-rad}, where each term is time-averaged between t=1000-1200 P$_{\mathrm{in}}$. Assuming a steady state ($\partial{E}/\partial t$=0), the energy flux divergence term on the left-hand side ($\bfnabla\cdot\left[(E+P)\bvp\right]$) should be balanced by three terms on the right-hand side: cooling from the disk radiation (-$Sr(E)$), heating from the stellar irradiation (-$\bfnabla\cdot \bF*$), and the work done by the stellar gravity ($\rho\mathbf{a_{\mathrm{grav}}} \cdot \bvp$). We can move the gravity term to the left-hand side and incorporate it into the energy flux divergence, considering it as the flux divergence of gravitational potential energy.

Figure \ref{fig:energy_budget} compares the disk cooling (left panel), stellar heating (middle panel), and the energy flux divergence (right panel). In the atmosphere, stellar heating is prominent, and most of the energy is radiated away by disk cooling, as both -$\bfnabla\cdot \bF_*$ and $Sr(E)$ exhibit similar values. However, in the midplane, stellar heating is nearly negligible, given our implementation of single-frequency stellar irradiation (Equation \ref{eq:IRRAD}), where few photons can penetrate below the $\tau_*=1$ surface. Interestingly, while disk cooling values are relatively small in the midplane, they still surpass stellar heating significantly. This additional disk cooling is offset by the negative energy flux divergence shown in the right panel.

Upon closer examination of each flux component, we find that the negative divergence arises from the advection of energy from the upper and lower atmospheres to the midplane. When we deactivate advection by freezing the hydrodynamics and solely conduct radiative transfer, the midplane temperature experiences a slight decrease by a few percent, indicating that energy advection can increase the temperature of the cold midplane by a few percent. We expect that when we include multi-wavelength stellar irradiation, the midplane can receive more stellar heating so that the influence of advection on the midplane temperature can be weaker.

\section{Disruption of the Inertial Wave}
Linear theory has unveiled the global model of VSI as an overstability, a destabilized inertial wave \citep{barker15}. Recently, \citet{svanberg22} (also see \citealt{stoll14}) used locally and vertically isothermal simulations to investigate the inertial wave patterns of VSI. A significant finding is that inertial waves associated with the corrugation mode could be identified in several radial wave zones separated at Lindblad resonances, each characterized by different frequencies. These wave zones also exhibit slightly different turbulence values.

\begin{figure*}
	\includegraphics[width=\linewidth]{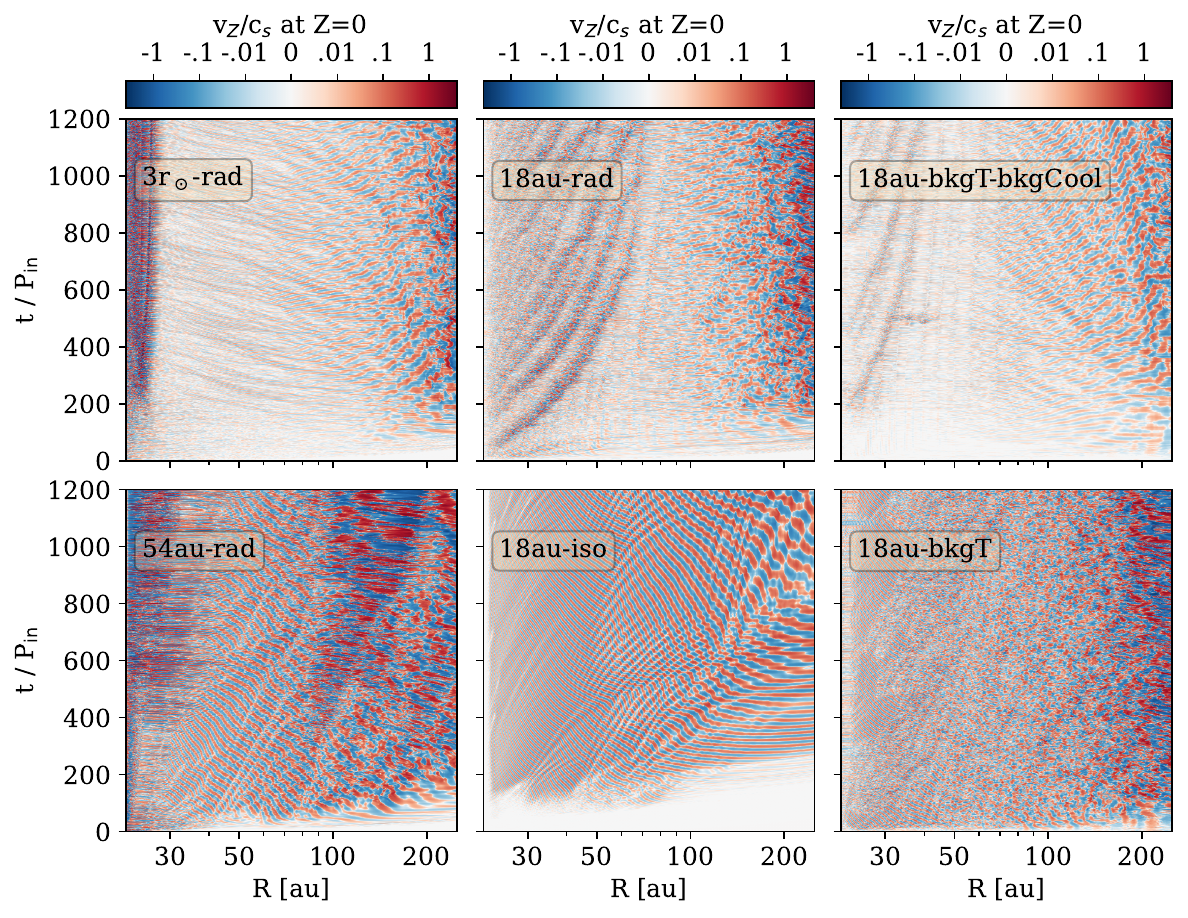}
    \caption{The time evolution of the vertical velocity (v$_{\mathrm{Z}}$) in the midplane from 0-1200 P$_\mathrm{in}$ for various models in the same layout as previous figures.}
    \label{fig:vZ_midplane_time_evolution}
\end{figure*}

\begin{figure*}
	\includegraphics[width=\linewidth]{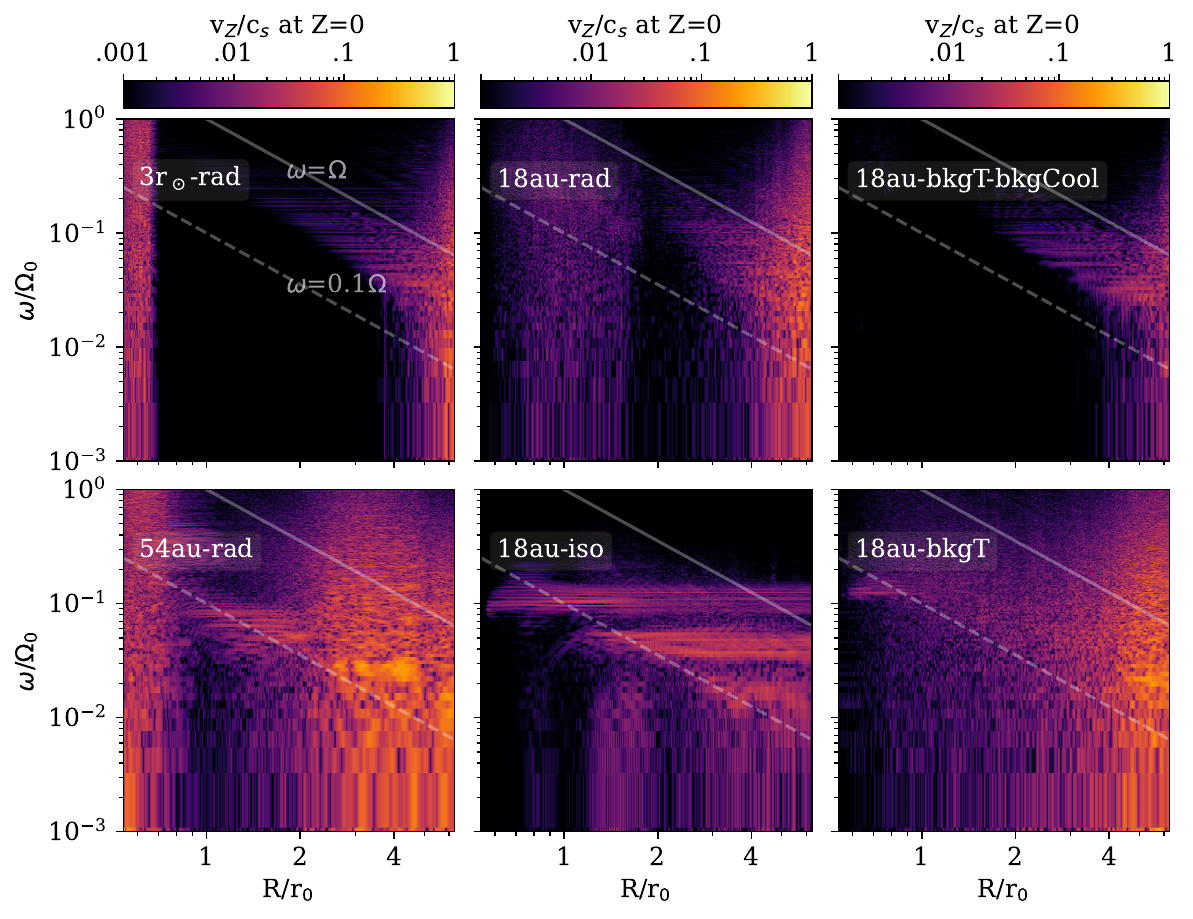}
    \caption{The Fourier transform of the time evolution of the vertical velocity in the midplane (Figure \ref{fig:vZ_midplane_time_evolution}). The frequency ($\omega$) is in unit of 1/$\Omega_{K,0}$, where $\Omega_{K,0}$ is the Keplerian frequency at 40 au. R is also normalized by r$_0$ = 40 au. The diagonal solid line indicates $\omega$ = $\Omega$ = $\Omega_{K,0} (R/r_0)^{-1.5}$, and the dashed line indicates $\omega$ = 0.1$\Omega$. The layout is the same as previous figures.}
    \label{fig:FFT_vZ}
\end{figure*}

\begin{figure*}
	\includegraphics[width=\linewidth]{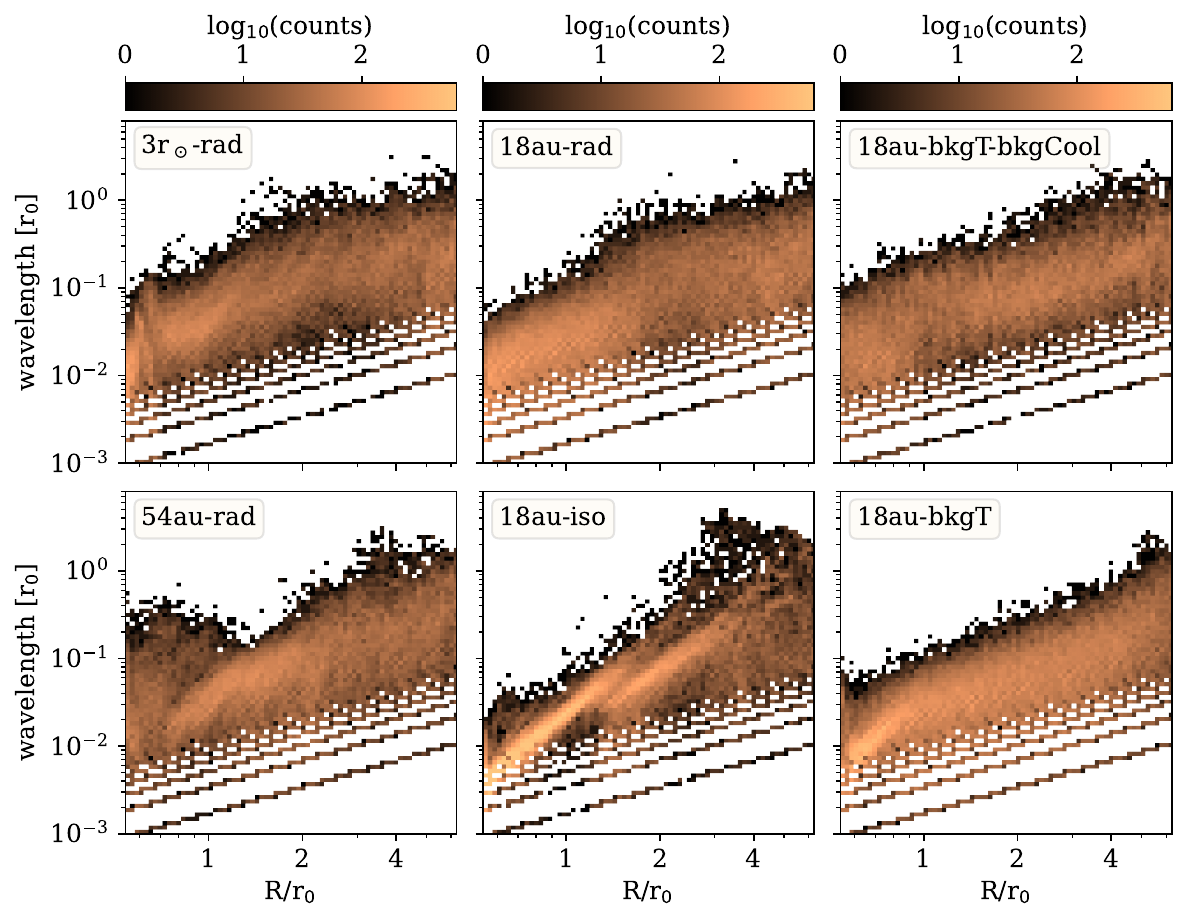}
    \caption{The wavelength occurrence of the vertical velocity (v$_Z$) in the midplane measured from every snapshot from t = 0-1200 P$_\mathrm{in}$, with 1 P$_\mathrm{in}$ as the interval. The wavelength is measured as the distance of the neighbouring zero crossing points. The brighter colors represent higher counts. The layout is the same as previous figures.}
    \label{fig:wavelengths}
\end{figure*}

However, when considering a self-consistent thermal structure that takes into account stellar irradiation, the inertial wave patterns become less distinct. Figure \ref{fig:vZ_midplane_time_evolution} illustrates the time evolution of the vertical velocity at the midplane, while the accompanying frequency and wavelength analyses are presented in Figures \ref{fig:FFT_vZ} and \ref{fig:wavelengths}. In the isothermal simulation \texttt{18au-iso}, we observe the classical corrugation model pattern as found in \citet{stoll14} and \citet{svanberg22}, characterized by alternating peaks and troughs moving radially. These patterns represent group velocity and phase velocities propagating in opposite directions (Figure 3 in \citealt{svanberg22}). In contrast, the remaining simulations do not exhibit this evident wave feature, except for \texttt{54au-rad} between 30-60 au, within the cavity and at the ring location. This is consistent with our observations in Figure \ref{fig:2DrhoTvZ}, where the corrugation mode is identified in $v_Z$ for \texttt{54au-rad} near the cavity. Similarly, the \texttt{18au-lowdens-rad} model (not displayed here) exhibits a wave pattern similar to the isothermal simulation. For the other models, we can still observe certain wave patterns, albeit different from the inertial wave patterns identified in the isothermal model.
The pronounced velocity in \texttt{3$r_\odot$-rad} is likely an inner boundary effect due to the disk surface's ($h/r$) discontinuity.
A comprehensive understanding of these wave patterns still needs ongoing studies.

In summary, the classical VSI inertial wave pattern only manifests when the disk is optically thin to stellar irradiation so that the disk is close to vertically isothermal and has a short cooling time. This can occur when the disk has low dust opacity (\texttt{18au-rad,lowdens}) or when the disk possesses a wide inner cavity (\texttt{54au-rad}).

\section{Asymmetry Above and Below the Midplane}
\label{sec:anticorrelation}
\begin{figure*}
	\includegraphics[width=\linewidth]{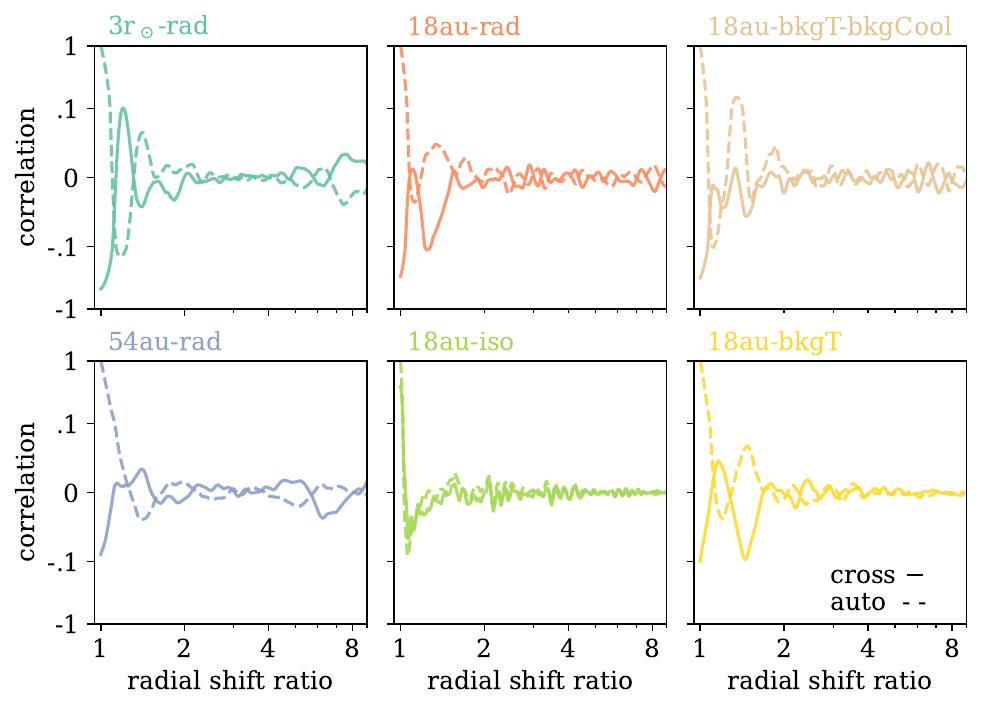}
    \caption{The cross-correlation functions of the v$_\mathrm{Z}$ between the upper and lower atmosphere shifted in the radial (r) direction with varying shift ratios, averaged between 0.2-0.25 radians above and below the midplane for t = 1000-1200 P$_{\mathrm{in}}$ (t = 500-700 P$_{\mathrm{in}}$ for \texttt{54au-rad}). The values are normalized by the auto-correlation functions in the upper atmosphere without a shift (shift ratio = 1). If the upper and lower atmosphere has the same v$_\mathrm{Z}$, i.e., v$_\mathrm{Z}$(r, $\theta$) = v$_\mathrm{Z}$(r, $\pi$ - $\theta$), the correlation should be unity at shift ratio = 1.}
    \label{fig:cross_correlation}
\end{figure*}

While we have demonstrated that the inertial wave pattern associated with the corrugation mode in classical VSI becomes less apparent in rad-hydro simulations or pure hydro simulations with self-consistent thermal structures, we can still observe some elongated stripes in the vertical direction for $v_Z$, as depicted in Figures \ref{fig:2DrhoTvZ} and \ref{fig:compareradbkg2D}. In contrast to isothermal simulations, the gas parcels above and below the midplane do not appear to move consistently in the same direction. The vertical velocity in \citet{melonfuksman23a} also shows anti-symmetry when the disk midplane is VSI-inactive ($f_\mathrm{dg}$=10$^{-4}$ therein), whereas corrugation mode only occurs when the VSI is active in the whole domain  ($f_\mathrm{dg}$=10$^{-3}$ therein). 

In Figure \ref{fig:cross_correlation}, we attempt to quantify these trends by measuring the autocorrelation function (dashed lines) for a horizontal cut within the range of 0.2-0.25 radians in the radial direction and a cross-correlation function (solid lines) between this cut and the one on the other side of the disk ($\pi-\theta$) at a specific snapshot and then averaging over 200 $\mathrm{P_{in}}$. Their values are normalized by the autocorrelation value with no radial shift (radial shift ratio = 1). Absolute values between 0.1 to 1 are presented in a log-scale, while absolute values between 0 and 0.1 are shown in a linear scale. The cross-correlation of the isothermal model (\texttt{18au-iso}) closely aligns with the autocorrelation function, suggesting that the upper and lower disks move in the same direction, echoing the n=1 corrugation model in Figure \ref{fig:compareradbkg2D}. In contrast, the remaining models exhibit trends where autocorrelation and cross-correlation functions have the opposite signs during the first few turnovers, indicating that the upper and lower disks are more likely to move in opposite directions. This anti-correlation between upper and lower surfaces resembles that of the n=2 breathing mode found in the initial growing phase of VSI \citep[e.g.,][]{nelson13, barker15}. However, this anti-correlation is not as strong as those in the linear growth phase, indicating that more than one modes are operating in this highly non-linear regime. A possible explanation is that the n=1 corrugation mode requires communication between the upper and lower disk. However, in our fiducial radiation models, the communication between two surfaces are disturbed by the temperature and cooling time stratification so that n=2 and other modes take over.

\section{Parameter List}
\begin{table}
	\centering
	\caption{Parameters used in the paper}
	\label{tab:parameters}
	\begin{tabular}{lr}
		\hline
		Symbol & Description \\
		\hline
		 r, $\mathrm{\theta}$, $\mathrm{\phi}$ & spherical polar coordinate\\
          R, Z, $\mathrm{\phi}$ & cylindrical coordinate \\
          $\mathrm{\pi}$/2 - $\mathrm{\theta}$ &  this value is 0 at the midplane \\
          $\mathrm{\Omega}$ &  orbital frequency\\ 
          $\mathrm{\Omega_K}$ &  Keplerian velocity\\
          r$_\odot$ & $\approx$ 0.0047 au, solar radius\\
          r$_0$ & = 40 au = 1 code unit of length\\
          r$_\mathrm{cav}$ & cavity size \\
          $\mathrm{\Omega_{K,0}^{-1}}$ & = 1 code unit of time $\approx$ 40 yr\\
          $p$   & midplane density power-law index \\
          $q$   & temperature power-law index \\
          $r$   & surface density power-law index \\
          $\gamma_g$ & = c$_p$/c$_V$ = 1.4, adiabatic index\\
          $R_\mathrm{ideal}$ = $k_b/m_p$ &  ideal gas constant \\
          $\mu$ & = 2.3 mean molecular weight\\
          $c_s$ & = ($R$T/$\mu$)$^{1/2}$, isothermal sound speed\\
          $P$ & gas pressure \\
          $\rho$ & gas density \\
          $\Sigma_\mathrm{g}$ &  gas surface density\\
          $\Sigma_\mathrm{g}$/$\Sigma_\mathrm{d}$ &  gas-to-dust mass radio\\
          v$_\mathrm{rms}$ & = root mean square velocity \citep[e.g.,][]{flock20}  \\
          v$_\mathrm{mag}$ & magnitude of the meridoinal velocity\\
          v$_0$ & unit velocity\\
          a$_0$ & characteristic sound speed\\
          T & temperature, assume T$_\mathrm{gas}$=T$_\mathrm{dust}$\\
          T$_0$ & = 1 code unit of temperature = 6.14$\times$10$^{3}$ K\\
          T$_\mathrm{floor}$ &= 6.14 K, temperature floor\\
          T$_\mathrm{ceiling}$ &= 614 K, temperature ceiling\\
          T$_\mathrm{\odot}$ & solar temperature\\
          r$_\mathrm{in}$ & = 21.6 au, inner boundary \\
          P$_\mathrm{in}$ & $\approx$ 100 yr, orbital period at r$_\mathrm{in}$  \\
          $\beta$ & dimensionless cooling time\\
          $\mathrm{\beta_c}$ & critical cooling time for VSI\\
          f$_\mathrm{s}$ & mass fraction of the small dust \\
          $h$ & gas vertical scale height \\
          $\mathrm{\tau_*}$ & stellar optical depth in the radial direction \\
          $\mathrm{\kappa_{P,d}}$, $\mathrm{\kappa_{R,d}}$ & Planck and Rosseland mean opacities (to dust)\\
          $\mathrm{\kappa_{P,g}}$, $\mathrm{\kappa_{R,g}}$ & Planck and Rosseland mean opacities (to gas)\\
          $\mathrm{\tau_{P,d}}$, $\mathrm{\tau_{R,d}}$ & disk optical depth in the vertical direction\\
          $M_\mathrm{acc}$ & mass accretion rate \\
          $\alpha_\mathrm{int}$ & vertically integrated $\alpha_{R,\phi}$ \\
          $\omega$ & angular frequency of the wave \\
          a & particle size \\
          St & Stokes number, or dimensionless stopping time\\
          
          T$_\mathrm{R,\phi}$, T$_\mathrm{Z,\phi}$ &  Reynolds stresses in R-$\phi$, Z-$\phi$ directions\\
          $\alpha_\mathrm{R}$, $\alpha_\mathrm{Z}$ &  = T$_\mathrm{R,\phi}$/<$P$>, T$_\mathrm{Z,\phi}$/<$P$>, turbulence levels\\
          
		\hline
	\end{tabular}
\end{table}

\end{document}